\documentclass[11pt,pre,superscriptaddress,showkeys,byrevtex]{revtex4}

\usepackage{times}
\usepackage{graphics}
\usepackage{amssymb,amsmath,amsthm}
\usepackage{amsfonts}

\newtheorem{defn}{Definition}[section]

\newtheorem{conj}[defn]{Conjecture}

\newtheorem{example}[defn]{Example}

\newtheorem{lemma}[defn]{Lemma}
\newtheorem{prop}[defn]{Proposition}

\newtheorem{remark}[defn]{Remark}
\newtheorem{thm}[defn]{Theorem}
\newtheorem{theorem}[defn]{Theorem}

\newcommand{\be}{\begin{equation}}
\newcommand{\ee}{\end{equation}}
\newcommand{\bea}{\begin{eqnarray}}
\newcommand{\eea}{\end{eqnarray}}
\newcommand{\beas}{\begin{eqnarray*}}
\newcommand{\eeas}{\end{eqnarray*}}
\newcommand{\brmk}{\begin{remark}\per\begin{em}}
\newcommand{\ermk}{\end{em}\end{remark}}
\newcommand{\goto}{\rightarrow}

\newcommand{\ink}{\rule{.5\baselineskip}{.55\baselineskip}}
\newcommand{\noi}{\noindent}
\newcommand{\lan}{\langle}
\newcommand{\ran}{\rangle}

\newcommand{\skp}{\vspace{\baselineskip}}

\newcommand{\per}{\hspace{-.072in}{\bf .  }}

\newcommand{\R}{{\mathbb R}}

\newcommand{\rsigma}{{\mathbb R}^{\sigma}}
\newcommand{\N}{\mathbb N}

\newcommand{\Z}{\mathbb Z}

\newcommand{\eu}{{\cal E}^u}
\newcommand{\ebeta}{{\cal E}_\beta}
\newcommand{\egbeta}{{\cal E}_{\beta,\gamma}}
\newcommand{\egammabeta}{{\cal E}_{\beta,\gamma}}
\newcommand{\X}{{\cal X}}

\newcommand{\thi}{\tilde{H}}

\newcommand{\mboxdomf}{\mbox{dom} \, f}

\newcommand{\mboxintdomf}{\mbox{int(dom} \, f)}

\newcommand{\mboxdoms}{\mbox{dom} \, s}
\newcommand{\mboxemdoms}{\mbox{{\em dom}} \, s}

\newcommand{\mboxintdoms}{\mbox{int(dom} \, s)}
\newcommand{\mboxemintdoms}{\mbox{{\em int(dom}} \, s)}
\newcommand{\cp}{\mathcal{P}}

\newcommand{\ts}{\textstyle}

\newcommand{\bexa}{\begin{example}\per\begin{em}}
\newcommand{\eexa}{\end{em}\end{example}}

\newcounter{bean}
\newcommand{\benuma}{\setlength{\labelwidth}{.25in}
\begin{list}%
{(\alph{bean})}{\usecounter{bean}}}
\newcommand{\eenuma}{\end{list}}

\def\theequation{\thesection.\arabic{equation}}
\def\theequation{\arabic{section}.\arabic{equation}}
\def\thedefn{\arabic{section}.\arabic{defn}}

\begin{document}

\title{Global Optimization, the Gaussian Ensemble, and Universal Ensemble
Equivalence\footnote{With great affection this paper is dedicated to Henry McKean on the occasionof his 75$^{\textrm{th}}$ birthday.}
}
\author{Marius Costeniuc}
\email{marius@mis.mpg.de}
\affiliation{\mbox{Max Planck Institute for Mathematics in the Sciences, Inselstrasse 22-26, D-04103 Leipzig, Germany}}

\author{Richard S.\ Ellis}
\email{rsellis@math.umass.edu}
\affiliation{Department of Mathematics and Statistics, University of Massachusetts, Amherst, MA 01003, USA}

\author{Hugo Touchette}
\email{htouchet@alum.mit.edu}
\affiliation{School of Mathematical Sciences, Queen Mary, University of London, London E1 4NS, UK}

\author{Bruce Turkington}
\email{turk@math.umass.edu}
\affiliation{Department of Mathematics and Statistics, University of Massachusetts, Amherst, MA 01003, USA}

\begin{abstract}
Given a constrained minimization problem, under what conditions does there exist a related, unconstrained problem having the same minimum points?  This basic question in global optimization motivates this paper, which answers it from the viewpoint of statistical mechanics.  In this context, it reduces to the fundamental question of the equivalence and nonequivalence of ensembles, which is analyzed using the theory of large deviations and the theory of convex functions.

In a 2000 paper appearing in the {\it Journal of Statistical Physics}, we gave necessary and sufficient conditions for ensemble equivalence and nonequivalence in terms of support and 
concavity properties of the microcanonical entropy.  
In later research we significantly extended those results by introducing a class of Gaussian ensembles, which are obtained from the  canonical ensemble by adding an exponential factor involving a quadratic function of the Hamiltonian. 
The present paper is an overview of our work on this topic. 
Our most important discovery is that even when the microcanonical and canonical ensembles are not equivalent, one can often find a Gaussian ensemble that satisfies a strong form of equivalence with the 
microcanonical ensemble known as universal equivalence. When translated back into optimization theory, this implies that an unconstrained minimization problem involving a Lagrange multiplier and a quadratic penalty function has the same minimum points as the original constrained problem.  

The results on ensemble equivalence discussed in this paper are illustrated in the context of the Curie-Weiss-Potts lattice-spin model.
\end{abstract}

\keywords{Equivalence of ensembles, Gaussian ensemble, 
microcanonical entropy, large deviation principle, Curie-Weiss-Potts model}

\maketitle

\section{Introduction}
\setcounter{equation}{0}
\label{section:intro}

At the beginning of his groundbreaking 1973 paper, Oscar Lanford describes
the underlying program of statistical mechanics \cite[p.\ 1]{Lan}.

\begin{quote}
The objective of statistical mechanics is to explain the macroscopic properties of matter on the basis of the behavior of the atoms and molecules of which it is composed. One of the most striking facts about macroscopic matter is that in spite of being fantastically complicated on the atomic level --- to specify the positions and velocities of all molecules in a glass of water would mean specifying something of the order of $10^{25}$ parameters --- its macroscopic behavior is describable in terms of a very small number of parameters; e.g., the temperature and density for a system containing only one kind of molecule. 
\end{quote}

Lanford shows how the theory of large deviations enables this objective to be realized.
In statistical mechanics 
one determines the macroscopic behavior of physical systems not from the 
deterministic laws of Newtonian mechanics, but from a probability distribution that expresses both the behavior of the system on the microscopic level and the intrinsic inability to describe precisely what is happening on that level.  
Using the theory of large deviations, one shows that, 
with probability converging to 1 exponentially fast as the number of particles tends to $\infty$,
the macroscopic behavior is describable in terms of a very small number of parameters 

The success of this program depends on the correct choice of probability distribution, also known as an ensemble.  One starts with a prior measure on configuration space, which, as an expression of the lack of information concerning the behavior of the system on the atomic level, is often taken to be the uniform measure. The most natural choice of ensemble is the microcanonical ensemble, obtained by conditioning the prior measure on the set of 
configurations for which the Hamiltonian per particle equals a constant energy $u$.  
Gibbs introduced a mathematically
more tractable probability distribution known as the Gibbs ensemble or the canonical ensemble, in which the conditioning that defines the microcanonical ensemble is replaced by an exponential factor 
involving the Hamiltonian and the inverse temperature
$\beta$, a parameter dual to the energy parameter $u$ \cite{Gibbs}.

Among other reasons, the 
canonical ensemble was introduced by Gibbs in the hope that 
in the limit $n \goto \infty$ the two
ensembles are equivalent; i.e., 
all macroscopic properties of the model obtained via the microcanonical ensemble
could be realized as macroscopic properties obtained via the canonical ensemble.  
While ensemble equivalence is valid for many standard and important models, 
ensemble equivalence does not hold in general, as numerous studies cited later in this introduction
show.  There are many examples of statistical mechanical models for which
nonequivalence of ensembles holds over a wide range of model
parameters and for which physically interesting microcanonical
equilibria are often omitted by the canonical ensemble.

\iffalse
In our research an analytic criterion for ensemble equivalence
was found in terms of concavity properties of the microcanonical 
entropy $s$ \cite{EHT1}.  If, for example, $s$ is strictly concave on its domain, then for any 
energy $u$ there exists a 
canonical ensemble such that macroscopic properties with respect to this ensemble coincide with macroscopic properties with respect to the microcanonical ensemble.  The two ensembles are said to be universally equivalent. If, on the other hand, $s$ is not concave on a subset $A$ of its domain, then for all energy values $u \in A$
the microcanonical ensemble is nonequivalent to all canonical ensembles for any choice of $\beta$.  In other words,
macroscopic properties with respect to the microcanonical ensemble are not macroscopic properties with respect to any canonical ensemble.  

In later research, summarized in \S XX, we showed that in the case of the nonequivalence of the canonical and microcanonical ensembles, universal equivalence often holds if one replaces the canonical ensemble by a Gaussian ensemble \cite{CosEllTouTur1}. The Gaussian ensemble is a perturbation of the canonical ensemble obtained by adding an exponential factor involving a quadratic function of the Hamiltonian.  
\fi

The present paper is an overview of our work on this topic.
One of the beautiful aspects of the theory is that it elucidates a fundamental issue in global optimization, which in fact motivated our work on the Gaussian ensemble.  Given a constrained minimization problem, 
under what conditions does there exist a related, unconstrained minimization problem having the same 
minimum points?  

In order to explain the connection between ensemble equivalence
and global optimization and in order to outline the contributions of this paper, we
introduce some notation.  Let ${\cal X}$ be a space,
$I$ a function mapping ${\cal X}$ into $[0,\infty]$, and $\tilde{H}$ a function
mapping ${\cal X}$ into $\R$.  
For $u \in \R$ we consider the following constrained minimization problem:
\be
\label{eqn:constrained}
\mbox{minimize } I(x) \mbox{ over } x \in {\cal X}
\mbox{ subject to the contraint } \tilde{H}(x) = u .
\ee
A partial answer to the 
question posed at the end of the preceding paragraph can be found by
introducing the following related, unconstrained minimization problem
for $\beta \in \R$:
\be
\label{eqn:unconstrained}
\mbox{minimize } I(x) + \beta \tilde{H}(x) 
\mbox{ over } x \in \X .
\ee
The theory of Lagrange multipliers outlines suitable 
conditions under which the solutions of the constrained problem
(\ref{eqn:constrained}) lie among the 
critical points of $I + \beta \tilde{H}$.  However, it does not give, as we will do in 
Theorems \ref{thm:equiv} and \ref{thm:main},
necessary and sufficient conditions for the solutions of (\ref{eqn:constrained}) to coincide with
the solutions of the unconstrained minimization problem (\ref{eqn:unconstrained}) and with the solutions
of the unconstrained minimization problem
appearing in (\ref{eqn:penalty2}).

We denote by $\eu$ and $\ebeta$ the respective sets of solutions of the minimization
problems (\ref{eqn:constrained}) and (\ref{eqn:unconstrained}).  
These problems arise in a natural way in 
the context of equilibrium statistical mechanics \cite{EHT1}, where
$u$ denotes the energy and $\beta$ the inverse temperature.  
As we will outline in Section 2, 
the theory of large deviations allows one to identify the solutions
of these problems as the respective sets of equilibrium macrostates for the microcanonical ensemble
and the canonical ensemble.  

The paper \cite{EHT1} analyzes equivalence of ensembles
in terms of relationships between $\eu$ and $\ebeta$.  
In turn, these relationships
are expressed in terms of support 
and concavity properties of the microcanonical entropy
\be
\label{eqn:intros}
s(u) = - \inf \{I(x) : x \in \X, \tilde{H}(x) = u\} .
\ee
The main results in \cite{EHT1} are summarized in Theorem \ref{thm:equiv}.   Part (a) of that theorem
states that if $s$ has a strictly supporting line 
at an energy value $u$, then full equivalence of ensembles holds in the sense that there
exists a $\beta$ such that $\eu = \ebeta$.  In particular,
if $s$ is strictly concave on $\mboxdoms$, then 
$s$ has a strictly supporting line at all $u$ 
except possibly boundary points
[Thm.\ \ref{thm:strictlyconcave}(a)] and thus
full equivalence of ensembles holds at all such $u$. 
In this case we say that the microcanonical
and canonical ensembles are universally equivalent.  

The most surprising result, given in part (c), 
is that if $s$ does not have a supporting line at $u$,
then nonequivalence of ensembles holds in the strong sense that
$\eu \cap \ebeta = \emptyset$ for all $\beta \in \rsigma$.  That is,
if $s$ does not have a supporting line at $u$ --- equivalently, 
if $s$ is not concave at $u$ --- then
microcanonical equilibrium macrostates cannot be realized canonically.
This is to be contrasted with part (d), which states
that for any $x \in \ebeta$ there exists $u$ such that
$x \in \eu$; i.e., canonical equilibrium macrostates can always be realized
microcanonically.  Thus of the two ensembles the microcanonical is the richer.

The paper \cite{CosEllTouTur1} addresses
the natural question suggested by part (c)
of Theorem \ref{thm:equiv}.  If the microcanonical ensemble is not equivalent with the
canonical ensemble on a subset of energy values $u$, then is it possible to replace the canonical ensemble
with another ensemble that is universally equivalent with the microcanonical ensemble?  We answered
this question by introducing a penalty function 
$\gamma[\tilde{H}(x) - u]^2$ into 
the unconstrained minimization problem 
(\ref{eqn:unconstrained}), obtaining the following:
\be
\label{eqn:penalty}
\mbox{minimize } I(x) + \beta \tilde{H}(x) + \gamma[\tilde{H}(x) - u]^2 
\mbox{ over } x \in \X .
\ee
Since for each $x \in \X$
\[
\lim_{\gamma \goto \infty} \gamma[\tilde{H}(x) - u]^2 
= \left\{ \begin{array}{ll} 0 & \ \mbox{ if } \thi(x) = u \\
\infty & \ \mbox{ if } \thi(x) \not = u,
\end{array}
\right.
\]
it is plausible that for all sufficiently large $\gamma$ minimum points of
the penalized problem (\ref{eqn:penalty}) are also
minimum points of the constrained problem (\ref{eqn:constrained}).  Since
$\beta$ can be adjusted, (\ref{eqn:penalty}) is equivalent to the following:
\be
\label{eqn:penalty2}
\mbox{minimize } I(x) + \beta \tilde{H}(x) + \gamma[\tilde{H}(x)]^2 
\mbox{ over } x \in \X .
\ee

The theory of large deviations allows one to identify the solution
of this problem as the set of equilibrium macrostates for the 
so-called Gaussian ensemble.  It is obtained from
the canonical ensemble by adding an exponential factor
involving $\gamma h_n^2$, where $h_n$ denotes 
the Hamiltonian energy per particle.  
The utility of the Gaussian ensemble rests on the simplicity with
which the quadratic function $\gamma u^2$ 
defining this ensemble
enters the formulation of ensemble equivalence.  
Essentially all the results in \cite{EHT1}
concerning ensemble equivalence, including
Theorem \ref{thm:equiv}, generalize
to the setting of the Gaussian ensemble by replacing
the microcanonical entropy $s(u)$ by the generalized microcanonical entropy 
\be
\label{eqn:sgamma}
s_\gamma(u) = s(u) - \gamma u^2 .
\ee
The generalization of Theorem \ref{thm:equiv}
is stated in Theorem \ref{thm:main},
which gives all possible relationships
between the set $\eu$ of equilibrium macrostates for the microcanonical ensemble
and the set $\egbeta$ of equilibrium macrostates for
the Gaussian ensemble.  These relationships
are expressed in terms of support and concavity properties of $s_\gamma$.

For the purpose of applications the most important consequence of Theorem \ref{thm:main}
is given in part (a), which states that if $s_\gamma$ has a strictly supporting line
at an energy value $u$, then full equivalence of ensembles holds in the sense that
there exists a $\beta$ such that $\eu = \egbeta$.  In particular, if $s_\gamma$ is strictly concave on 
$\mboxdoms$, then $s_\gamma$
has a strictly supporting line at all $u$ 
except possibly boundary points
[Thm.\ \ref{thm:sgstrictlyconcave}(a)] and thus full equivalence
of ensembles holds at all such $u$.  In this case we say that the microcanonical and
Gaussian ensembles are universally equivalent.

In the case in which $s$ is $C^2$ and $s''$ is
bounded above on the interior of $\mboxdoms$,
then the strict concavity of $s_\gamma$ is
easy to show.  In fact, the strict concavity
is a consequence of
\[
s_\gamma''(u) = s''(u) - 2 \gamma < 0 \ \mbox{ for all } u \in \mboxintdoms ,
\]
and this in turn is valid for all sufficiently
large $\gamma$ [Thm.\ \ref{thm:applysigma1}].  For such $\gamma$ it follows,
therefore, that the microcanonical and Gaussian ensembles are
universally equivalent.

Defined in (\ref{eqn:gencanon}), the Gaussian ensemble is mathematically much more 
tractable than the microcanonical ensemble, which is defined in 
terms of conditioning. The simpler form of the Gaussian ensemble 
is reflected in the simpler form of the unconstrained minimization 
problem (\ref{eqn:penalty2}) defining the set $\egbeta$ of Gaussian equilibrium 
macrostates. In (\ref{eqn:penalty2}) the constraint appearing in the minimization 
problem (\ref{eqn:constrained}) defining the set $\eu$ of microcanonical equilibrium 
macrostates is replaced by the linear and quadratic terms involving 
$\tilde{H}(x)$. The virtue of the Gaussian formulation should be clear. 
When the microcanonical and Gaussian ensembles are universally equivalent, 
then from a numerical point of view, it is
better to use the Gaussian ensemble because this ensemble, contrary
to the microcanonical one, does not involve an equality constraint,
which is difficult to implement numerically.  Furthermore, within
the context of the Gaussian ensemble, it is possible to use Monte
Carlo techniques without any constraint on the sampling \cite{CH1,CH2}.

By giving necessary and sufficient conditions
for the equivalence of the three ensembles in Theorems \ref{thm:equiv}
and \ref{thm:main}, we make contact
with the duality theory of global optimization and the method of
augmented Lagrangians \cite[\S2.2]{Bertsekas}, \cite[\S6.4]{Minoux}.
In the context of global optimization the primal function and the dual
function play the same roles that the microcanonical
entropy (resp., generalized microcanonical entropy)
and the canonical free energy (resp., Gaussian 
free energy) play in
statistical mechanics.  Similarly, the replacement of the Lagrangian
by the augmented Lagrangian in global optimization is paralleled by
our replacement of the canonical ensemble by the Gaussian ensemble.

The Gaussian ensemble is a special case of the generalized canonical ensemble,
which is obtained from the canonical ensemble by adding an exponential
factor involving $g(h_n)$, where $g$ is a continuous function that
is bounded below.  Our paper \cite{CosEllTouTur1} gives all possible 
relationships between the sets of equilibrium macrostates for the 
microcanonical and generalized canonical ensembles in terms of support
and concavity properties of an appropriate entropy function.  Our paper
\cite{TouCosEllTur} shows that the generalized canonical ensemble 
can be used to transform metastable or unstable nonequilibrium macrostates
for the standard canonical ensemble into stable equilibrium macrostates 
for the generalized canonical ensemble.

Equivalence and nonequivalence of ensembles is the subject of a large literature. An overview
is given in the introduction of \cite{LewPfiSul2}.  A number of theoretical papers on this topic, including 
\cite{DeuStrZes,EHT1,EyiSpo,Geo,LewPfiSul1,LewPfiSul2,RoeZes}, investigate
equivalence of ensembles using the theory of large deviations.  
In \cite[\S7]{LewPfiSul1} and \cite[\S7.3]{LewPfiSul2}   
there is a discussion of  nonequivalence of ensembles for the 
simplest mean-field model in statistical mechanics; namely, the
Curie-Weiss model of a ferromagnet.  
However, despite the mathematical
sophistication of these and other studies, none of them except
for our papers \cite{CosEllTouTur1,EHT1} explicitly addresses
the general issue of the nonequivalence of ensembles.

Nonequivalence of ensembles  
has been observed in a wide range of systems that involve 
long-range interactions and that can be studied by the methods of 
\cite{CosEllTouTur1,EHT1}.  In all of these cases the microcanonical formulation gives
rise to a richer set of equilibrium macrostates. 
\iffalse
Interestingly, this
nonequivalence is found to occur in many applications of the
equilibrium theory, especially in the negative temperature regimes of
the 
\fi
For example,
it has been shown computationally that the strongly reversing zonal-jet
structures on Jupiter as well as the Great Red Spot fall into the
nonequivalent range of an appropriate microcanonical ensemble \cite{turmajhavdib}.
Other models for which ensemble nonequivalence has been observed include
a number of long-range, mean-field spin models
including the Hamiltonian mean-field model \cite{DauLatRapRufTor,LatRapTsa2}, 
the mean-field X-Y model \cite{DHR}, 
and the mean-field Blume-Emery-Griffith model \cite{BMR1,BMR2,ETT}.  
For a mean-field version of the 
Potts model called the Curie-Weiss-Potts model, equivalence and nonequivalence of ensembles 
is analyzed in detail in \cite{CosEllTou1,CosEllTou2}.
Ensemble nonequivalence has also been observed in models of 
turbulent vorticity dynamics
\cite{CLMP,DibMajGro,DibMajTur,EHT2,EyiSpo,KieLeb,RobSom}, models of plasmas
\cite{KieNeu2,SmiOne}, gravitational systems \cite{Gross1,HerThi,LynBelWoo,Thi2},
and a model of the Lennard-Jones gas \cite{BorTsa}.
A detailed discussion of ensemble nonequivalence for models of 
coherent structures in two dimensional turbulence is given
in \cite[\S1.4]{EHT1}.

Gaussian ensembles were introduced in \cite{Heth1987} and studied further in 
\cite{CH1,CH2,Stump1987,JPV,Stump21987}. As these papers discuss, an important feature 
of Gaussian ensembles is that they 
allow one to account for ensemble-dependent effects in finite systems.
Although not referred to by name, the Gaussian ensemble 
also plays a key role in \cite{KieLeb}, where it is used to address
equivalence-of-ensemble questions for a point-vortex model of fluid
turbulence.  

Another seed out of which the research summarized in the
present paper germinated is the paper
\cite{EHT2}.  There we study the equivalence of the microcanonical
and canonical ensembles for statistical equilibrium models of coherent
structures in two-dimensional and quasi-geostrophic turbulence.
Numerical computations demonstrate that, as in other cases,
nonequivalence of ensembles occurs
over a wide range of model parameters and that physically interesting
microcanonical equilibria are often omitted by 
the canonical ensemble.  In addition, in 
Section 5 of \cite{EHT2}, we establish the nonlinear stability of the steady
mean flows corresponding to microcanonical equilibria 
via a new Lyapunov argument.  The associated stability theorem refines the
well-known Arnold stability theorems, which do not apply when the
microcanonical and canonical ensembles are not equivalent.  The Lyapunov
functional appearing in this new stability 
theorem is defined in terms of a generalized thermodynamic potential similar in form to 
$I(x) + \beta \thi(x) + \gamma [\thi(x)]^2$, 
the minimum points of which define the set of equilibrium macrostates
for the Gaussian ensemble [see (\ref{eqn:egbeta})].  
\iffalse
Such Lyapunov functionals arise
in the study of constrained optimization problems, where they are known as
augmented Lagrangians \cite{Bertsekas,Minoux}.  
\fi

Our goal in this paper is to give an overview of our theoretical work on ensemble
equivalence presented in \cite{CosEllTouTur1,EHT1}.  The paper \cite{CosEllTouTur2}
investigates the physical principles underlying this theory.
In Section 2 of the present paper, we first state the hypotheses on the statistical mechanical models to which the theory of the present paper applies.  We then define the three ensembles ---
microcanonical, canonical, and Gaussian --- and specify the three associated sets of equilibrium macrostates in terms of large deviation principles.  In Section 3 we state two sets of results on ensemble equivalence.
The first involves the equivalence of the microcanonical and canonical ensembles, necessary and sufficient conditions for which are given in terms of support properties of the microcanonical entropy $s$
defined in (\ref{eqn:intros}).  The second involves the equivalence of the microcanonical and Gaussian ensembles, necessary and sufficient conditions for which are given in terms of support properties of the generalized microcanonical entropy $s_\gamma$ 
defined in (\ref{eqn:sgamma}).  Section 4 addresses a basic foundational issue in statistical
mechanics.  There we show that when the canonical ensemble is nonequivalent to the
microcanonical ensemble on a subset of energy
values $u$, it can often be replaced by a Gaussian ensemble that is 
universally equivalent to the microcanonical ensemble.  In Section 5 the results on ensemble equivalence discussed in this paper are illustrated in the context of the Curie-Weiss-Potts lattice-spin model, a mean-field approximation
to the nearest-neighbor Potts model.  Several of the results presented near the end of this section are new.

\section{Definitions of Models and Ensembles}
\setcounter{equation}{0}
\label{section:definitions}

\renewcommand{\theequation}{\arabic{section}.\arabic{equation}}
\renewcommand{\thedefn}{\arabic{section}.\arabic{defn}}

One of the objectives of this paper is to show that when the canonical ensemble
is nonequivalent to the microcanonical ensemble on a subset of energy values $u$, 
it can often be replaced by a Gaussian ensemble that is equivalent
to the microcanonical ensemble for all $u$.  Before introducing the various
ensembles as well as the methodology for proving this result, we first 
specify the class of statistical mechanical models under consideration.
The models are defined in terms of the following quantities. 

\begin{enumerate}
\item A sequence of probability spaces $(\Omega_n, {\cal F}_n,P_n)$
indexed by $n \in \N$, which typically represents a sequence of finite dimensional systems.  
The $\Omega_n$ are the configuration spaces, 
$\omega \in \Omega_n$ are the microstates, and the $P_n$ are the prior measures 
on the $\sigma$ fields ${\cal F}_n$.

\item A sequence of positive scaling constant $a_n \goto \infty$ as $n \goto \infty$.  
In general $a_n$ equals the total number of degrees of freedom in the model.
In many cases $a_n$ equals the number of particles.

\item For each $n \in \N$ a measurable
functions $H_{n}$ mapping $\Omega_n$ into $\R$.  
For $\omega \in \Omega_n$
we define the energy per degree of freedom by
\[
h_{n}(\omega) = \frac{1}{a_n} H_{n}(\omega) .
\] 
\end{enumerate}

\noi
Typically, 
$H_n$ in item 3 equals
the Hamiltonian, which is associated with energy conservation in the model. 
The theory is easily generalized by replacing $H_n$ by a vector of appropriate functions
representing additional dynamical invariants associated with the model
\cite{CosEllTouTur1,EHT1}.

A large deviation analysis of the general model is possible provided 
that there exist a space of macrostates, 
macroscopic variables, and an interaction representation function and provided 
that the macroscopic variables satisfy the large deviation principle
(LDP) on the space of macrostates.  These concepts are explained next.

\begin{enumerate}

\item[4.] {\bf Space of macrostates}.  This is a complete, separable metric space ${\cal X}$,
which represents the set of all possible macrostates.

\item[5.] {\bf Macroscopic variables.}  These are a sequence of random variables
$Y_n$ mapping $\Omega_n$ into ${\cal X}$.  These functions associate a 
macrostate in $\X$ with each microstate $\omega \in \Omega_n$.

\item[6.] {\bf Interaction representation function.}  This is a
bounded, continuous functions $\tilde{H}$ 
mapping ${\cal X}$ into $\R$ such that as $n \rightarrow \infty$
\be 
\label{eqn:interaction}
h_{n}(\omega)=\tilde{H}(Y_n(\omega)) + \mbox{o}(1) \ \ \mbox{ uniformly for }
\omega \in \Omega_n ;
\ee
i.e., 
\[
\lim_{n  \rightarrow \infty} \sup_{\omega \in \Omega_n} 
|h_{n}(\omega)-\tilde{H}(Y_n(\omega))| = 0 .
\]  
The function $\tilde{H}$ 
enable us to write $h_{n}$, either exactly or asymptotically,
as a function of the macrostate via the macroscopic variables $Y_n$.  

\item[7.] {\bf LDP for the macroscopic variables.}  There exists a 
function $I$ mapping ${\cal X}$ into $[0, \infty]$ and having
compact level sets such that with respect to
$P_n$ the sequence $Y_n$ satisfies the LDP on ${\cal X}$ with rate function $I$
and scaling constants $a_n$.
In other words, for any closed subset $F$ of $\X$ 
\[
\limsup_{n \goto \infty} \frac{1}{a_n} \log P_n\{Y_n \in F\} \leq - \inf_{x \in F}I(x) ,
\]
and for any open subset $G$ of $\X$ 
\[
\liminf_{n \goto \infty} \frac{1}{a_n} \log P_n\{Y_n \in G\} \geq - \inf_{x \in G}I(x) .
\]
It is helpful to summarize the LDP by the formal notation
$P_n\{Y_n \in dx\} \asymp \exp[-a_n I(x)]$.   This notation expresses the fact
that, to a first degree of approximation, $P_n\{Y_n \in dx\}$ behaves like an
exponential that decays to 0 whenever $I(x) > 0$.
\end{enumerate}

A wide variety of statistical mechanical models satisfy the hypotheses
listed in items 1--7 at the start of this section
and so can be studied by the methods of \cite{CosEllTouTur1,EHT1}.
These include the following.
\begin{enumerate}
\item The mean-field Blume-Emery-Griffiths model \cite{BEG} is
one of the simplest lattice-spin
models known to exhibit, in the mean-field approximation, both a continuous, second-order phase transition
and a discontinuous, first-order phase transition.   
The space of macrostates for this model is the set of probability measures on a certain finite set,
the macroscopic variables are the empirical measures associated with the spin configurations, and the associated 
LDP is Sanov's Theorem, for which the rate function is a relative entropy.
Various features of this model are studied in \cite{BMR1,BMR2,EllOttTou,ETT}.
\item
The Curie-Weiss-Potts model is a
mean-field approximation to the nearest-neighbor Potts model \cite{Wu}. 
For the Curie-Weiss-Potts model, the space of macrostates, the macroscopic variables,
and the associated LDP are similar to those in the mean-field Blume-Emery-Griffiths model. 
The Curie-Weiss-Potts model nicely illustrates the general results on ensemble
equivalence discussed in this paper and is discussed in Section \ref{section:cwp}.
\item Short-range spin systems such as the Ising model on $\Z^d$
and numerous generalizations can also be handled by the methods of this paper.  The
large deviation techniques required to analyze these models
are much more subtle than in the case of 
the long-range, mean-field models considered in items 1 and 2.  
For the Ising model the space of macrostates 
is the space of translation-invariant probability measures on $\Z^d$, the macroscopic
variables are the empirical processes associated with the spin configurations, 
and the rate function in the associated LDP is the
mean relative entropy \cite{Ell,FoeOre,Olla}.
\item The Miller-Robert model is a model of coherent structures in an ideal,
two-dimensional fluid that includes all the exact invariants of the
vorticity transport equation \cite{Miller,Robert}. The space of macrostates is the space of Young measures
on the vorticity field.  The large deviation analysis of this model developed first in \cite{Robert}
and more recently in \cite{BouEllTur} gives a rigorous derivation of maximum entropy principles
governing the equilibrium behavior of the ideal fluid.
\item  In geophysical applications, another
version of the model in item 4 is preferred, in which the enstrophy integrals
are treated canonically and the energy and circulation are treated
microcanonically \cite{EHT2}.  In those formulations, the space of
macrostates is $L^2(\Lambda)$ or $L^{\infty}(\Lambda)$ depending on
the contraints on the voriticty field.  The large deviation analysis
is carried out in \cite{EHT3}.  The paper
\cite{EHT2} shows how the nonlinear stability of the steady mean flows arising as
equilibrium macrostates can be established by
utilizing the appropriate generalized thermodynamic potentials.
\item A statistical
equilibrium model of solitary wave structures in dispersive wave
turbulence governed by a nonlinear Schr\"odinger equation is studied in
\cite{EllJorOttTur}.  The large deviation analysis
given in \cite{EllJorOttTur} 
derives rigorously the concentration phenomenon observed in long-time
numerical simulations and predicted by mean-field approximations
\cite{JorTurZir,LebRosSpe2}.  The space of macrostates
is $L^2(\Lambda)$, where $\Lambda$ is a bounded interval or more
generally a bounded domain in $\R^d$.  The macroscopic variables
are certain Gaussian processes.
\end{enumerate}

We now return to the general theory, first
introducing the function whose support and concavity properties 
completely determine all aspects of ensemble equivalence and nonequivalence.
This function is the microcanonical entropy, defined for $u \in \R$ by
\be 
\label{eqn:entropy}
s(u) = -\inf \{I(x) : x \in {\cal X}, \tilde{H}(x) = u\} .
\ee
Since $I$ maps ${\cal X}$ into $[0,\infty]$, $s$ maps
$\rsigma$ into $[-\infty,0]$.  Moreover, since $I$ is lower semicontinuous 
and $\thi$ is continuous on $\X$, $s$ is 
upper semicontinuous on $\rsigma$. 
We define $\mbox{dom} \, s$ to be the set of $u \in \rsigma$ for which $s(u) > -\infty$.
In general, $\mboxdoms$ is nonempty since $-s$ is a rate function
\cite[Prop.\ 3.1(a)]{EHT1}.  
For each $u \in \mbox{dom} \, s$, $r > 0$,
$n \in \N$, and set $B \in {\cal F}_n$ the microcanonical ensemble 
is defined to be the conditioned measure
\be
\label{eqn:microens}
P^{u,r}_n \{B\}=P_n \{B \mid h_n \in [u-r,u+r]\} .
\ee
As shown in \cite[p.\ 1027]{EHT1},
if $u \in \mbox{dom} \, s$, then for all sufficiently large $n$,
$\, P_n\{h_n \in [u-r,u+r]\} > 0$; thus the
conditioned measures  $P^{u,r}_n $ are well defined.

A mathematically more tractable probability measure is
the canonical ensemble.               
For each $n \in \N$, $\beta
\in \R$, and set $B \in {\cal F}_n$ we define the partition function
\[ 
Z_n(\beta) = \int_{\Omega_n}  \exp [-a_n \beta h_n] \, dP_n ,
\]
which is well defined and finite, and the probability measure
\be
\label{eqn:canonens}
P_{n, \beta}\{B\} = \frac{1}{Z_n(\beta)} \cdot \int_{B} \exp [-a_n \beta h_n] \, dP_n .
\ee
The measures $P_{n, \beta}$ are Gibbs states that define the canonical
ensemble for the given model.  

The Gaussian ensemble is a natural perturbation of the
canonical ensemble.  For each $n \in \N$,
$\beta \in \R$, and $\gamma \in [0,\infty)$ we define
the Gaussian partition function
\be
Z_{n}(\beta,\gamma)=\int_{\Omega_n}  \exp [-a_n \beta h_n
-a_n \gamma h_n^2] \, dP_n .
\ee
This is well defined and finite because the $h_n$ are bounded.  
For $B \in {\cal F}_n$ we
also define the probability measure
\be
\label{eqn:gencanon} 
P_{n, \beta,\gamma}\{B\} = \frac{1}{Z_{n}(\beta,\gamma)} \cdot \int_{B}
\exp [-a_n \beta h_n - a_n \gamma h_n^2] \, dP_n ,
\ee
which we call the Gaussian canonical ensemble.
One can generalize this by replacing the quadratic function
by a continuous function $g$ that is bounded
below.  This gives rise to the generalized canonical ensemble,
which the theory developed in \cite{CosEllTouTur1} allows one to treat. 

Using the theory of large deviations, one introduces the sets of equilibrium macrostates
for each ensemble.  
It is proved in \cite[Thm.\ 3.2]{EHT1} 
that with respect to the microcanonical ensemble
$P_n^{u,r},$ $Y_n$ satisfies the LDP on ${\cal X}$, in the
double limit $n \rightarrow \infty$ and $r \rightarrow 0$, with rate
function
\be
\label{eqn:iu}
I^u(x) = \left\{
\begin{array}{ll}
           I(x)+s(u) & \ \mbox{ if } \: \tilde{H}(x)=u
\\
           \infty & \ \mbox{ otherwise }  . 
\end{array}
\right.
\ee
$I^u$ is nonnegative on $\X$, and for $u \in \mboxdoms$, $I^u$ 
attains its infimum of 0 on the set
\bea
\label{eqn:eu}
 {{\cal E}}^u & = & \{x \in {\cal X} : I^u(x)=0\} \\
& = & \{x \in \X : I(x) \mbox{ is minimized subject to } \thi(x) = u\} .
\nonumber 
\eea
This set is precisely the set of solutions of the constrained minimization
problem (\ref{eqn:constrained}).

In order to state the LDPs for the other two ensembles, we 
bring in the canonical free energy, defined for $\beta \in \R$ by 
\[
\varphi(\beta) = - \lim_{n \goto \infty} \frac{1}{a_n}
\log Z_{n}(\beta) ,
\]
and the Gaussian free energy, defined for $\beta \in \R$ and $\gamma \geq 0$ by 
\[
\varphi(\beta,\gamma) = - \lim_{n \goto \infty} \frac{1}{a_n}
\log Z_{n}(\beta,\gamma) .
\]
It is proved in \cite[Thm.\ 2.4]{EHT1} that the limit defining $\varphi(\beta)$ exists
and is given by
\be
\label{eqn:varphi}
\varphi(\beta) =  \inf_{y \in {\cal X}} \{I(y) + \beta \tilde{H}(y)\}
\ee
and that with respect to $P_{n,\beta}$, $Y_n$ satisfies
the LDP on ${\cal X}$ with rate function
\be
\label{eqn:ibeta}
I_\beta(x)=I(x)+ \beta \tilde{H}(x) - \varphi(\beta) .
\ee
$I_\beta$ is nonnegative on $\X$ and attains its infimum of 0 on
the set
\bea
\label{eqn:ebeta}
{{\cal E}}_\beta & =& \{x \in {\cal X} : I_\beta(x)=0\} \\
& = & \{x \in \X : I(x) + \lan \beta,\thi(x) \ran \mbox{ is minimized}\} . \nonumber
\eea
This set is precisely the set of solutions of the unconstrained minimization
problem (\ref{eqn:unconstrained}).

A straightforward extension of these results shows
that the limit defining $\varphi(\beta,\gamma)$ exists and is given by
\be
\label{eqn:varphig}
\varphi(\beta,\gamma) = \inf_{y \in \X}\{I(y) + \beta \thi(y) + \gamma [\thi(y)]^2\}
\ee
and that with respect to $P_{n,\beta,g}$, $Y_n$ satisfies
the LDP on ${\cal X}$ with rate function
\be
\label{eqn:ibetag}
I_{\beta,\gamma}(x)=I(x)+ \beta \tilde{H}(x) + \gamma[\thi(x)]^2 - \varphi(\beta,\gamma) .
\ee
$I_{\beta,\gamma}$ is nonnegative on $\X$ and attains its infimum of 0 on
the set
\bea
\label{eqn:egbeta}
{{\cal E}}_{\beta,\gamma} & = & \{x \in {\cal X} : I_{\beta,\gamma}(x)=0\} \\
& = &  \{x \in \X : I(x) + \lan \beta,\thi(x) \ran + \gamma[\thi(x)]^2 
\mbox{ is minimized}\} . \nonumber
\eea
This set is precisely the set of solutions of the penalized minimization
problem (\ref{eqn:penalty2}).

For $u \in \mboxdoms$, let $x$ be any element of $\X$ satisfying
$I^u(x) > 0$.  The formal notation
\[
P_{n}^{u,r}\{Y_n \in dx\} \asymp e^{-a_n I^u(x)}
\]
suggests that $x$ has an exponentially small probability of
being observed in the limit  $n \goto \infty$, $r \goto 0$.  
Hence it makes sense to identify $\eu$
with the set of microcanonical equilibrium macrostates.  In the same way we identify
with $\ebeta$ the set of canonical equilibrium macrostates and with 
$\egbeta$ the set of generalized canonical equilibrium macrostates.
A rigorous justification is given in \cite[Thm.\ 2.4(d)]{EHT1}.

\section{Equivalence and Nonequivalence of the Three Ensembles}
\setcounter{equation}{0}
\label{section:equivnonequiv}

Having defined the sets 
of equilibrium macrostates $\eu$, $\ebeta$, and $\egbeta$ for the microcanonical,
canonical and Gaussian ensembles, we now show how these sets are related to one
another.  In Theorem \ref{thm:equiv} we state the results proved
in \cite{EHT1} concerning equivalence and nonequivalence of the microcanonical
and canonical ensembles.  Then in Theorem \ref{thm:main} 
we extend these results to the Gaussian ensemble \cite{CosEllTouTur1}.

Parts (a)--(c) of Theorem \ref{thm:equiv}
give necessary and sufficient conditions, in terms of 
support properties of $s$, 
for equivalence and nonequivalence of $\eu$ and $\ebeta$.
These assertions are proved in Theorems 4.4 and 4.8 in \cite{EHT1}.
Part (a) states that $s$ has a 
strictly supporting line at $u$ if and only if full equivalence of ensembles
holds; i.e., if and only if there exists a $\beta$ such that
$\eu = \ebeta$.  The most surprising result, given in part (c), is 
that $s$ has no supporting line at $u$ if and only if nonequivalence of ensembles
holds in the strong sense that $\eu \cap \ebeta = \emptyset$ for 
all $\beta$.  
Part (c) is to be contrasted with part (d), which
states that for any $\beta$ 
canonical equilibrium macrostates can always be realized
microcanonically.  Part (d) is proved in Theorem 4.6 in \cite{EHT1}.
Thus one conclusion of this theorem is that at the level 
of equilibrium macrostates the microcanonical 
ensemble is the richer of the two ensembles.  

\begin{theorem}  In parts {\em (a)}, {\em (b)}, and {\em (c)}, 
$u$ denotes any point in $\mbox{{\em dom}} \, s$.
\label{thm:equiv}  

{\em (a)}  \mbox{{\em {\bf Full equivalence.}}}  \ There exists $\beta$
such that ${\cal E}^u = {\cal E}_\beta$ 
if and only if $s$ has a strictly supporting line at $u$ with slope $\beta$; i.e., 
\[
s(v) < s(u) + \beta (v-u) \mbox{ for all } v \not = u \,.
\]

{\em (b)} \mbox{{\em {\bf Partial equivalence.}}} \ There exists $\beta$
such that
${\cal E}^u \subset {\cal E}_\beta$ but ${\cal E}^u \not = {\cal E}_\beta$
if and only if $s$ has a nonstrictly supporting line at $u$ with slope $\beta$; i.e., 
\[
s(v) \leq  s(u) + \beta (v-u) \mbox{ for all } v 
\mbox{ with equality for some } v \not = u .
\]

{\em (c)} \mbox{{\em {\bf Nonequivalence.}}} \ For all $\beta$,
${\cal E}^u \cap {\cal E}_\beta = \emptyset$
if and only if $s$ has no supporting line at $u$; i.e., 
\[
\mbox{for all } \beta  \mbox{ there exists } v \mbox{ such that }
s(v) > s(u) + \beta (v-u) .
\]  
\iffalse
Except possibly boundary points of $\mboxemdoms$,
the latter condition is equivalent to the nonconcavity of $s$ at $u$. 
\fi

{\em (d)}  \mbox{{\em {\bf Canonical is always realized microcanonically.}}}
\ For any $\beta \in \R$ we have $\tilde{H}({\cal E}_\beta) \subset 
\mbox{\em dom} \, s$ and
\[
{\cal E}_\beta = \bigcup_{u \in \tilde{H}({\cal E}_\beta)} {\cal E}^u .
\]
\end{theorem}

We highlight several features of the theorem in order to illuminate
their physical content.  In part (a) we assume that for a given $u \in \mboxdoms$
there exists a unique $\beta$ such that $\eu = \ebeta$.  
If $s$ is differentiable at $u$ and $s$ and the double-Legendre-Fenchel transform $s^{**}$
are equal in a neighborhood of $u$, then $\beta$ is given by the 
standard thermodynamic formula $\beta = s'(u)$ 
\cite[Thm.\ A.4(b)]{CosEllTouTur1}.  The inverse
relationship can be obtained from part (d) of the theorem under
the assumption that $\ebeta$ consists of a unique macrostate
or more generally that for all $x \in \ebeta$ the values
$\tilde{H}(x)$ are equal.  Then $\ebeta = {\cal E}^{u(\beta)}$, where
$u(\beta) = \tilde{H}(x)$ for any $x \in \ebeta$; $u(\beta)$
denotes the mean energy realized at equilibrium in the canonical ensemble.  The relationship
$u = u(\beta)$ inverts the relationship $\beta = s'(u)$.  Partial ensemble
equivalence can be seen in part (d) under the assumption that for a given $\beta$,
$\ebeta$ can be partitioned into at least two sets ${\cal E}_{\beta,i}$ such that 
for all $x \in {\cal E}_{\beta,i}$ the values $\tilde{H}(x)$ are equal but 
$\tilde{H}(x) \not = \tilde{H}(y)$
whenever $x \in {\cal E}_{\beta,i}$ and $y \in {\cal E}_{\beta,j}$ for $i \not = j$. Then
$\ebeta = \bigcup_{i}{\cal E}^{u_i(\beta)}$, where $u_i(\beta) = \tilde{H}(x)$, $x \in {\cal E}_{\beta,i}$.
Clearly, for each $i$, ${\cal E}^{u_i(\beta)} \subset \ebeta$ but ${\cal E}^{u_i(\beta)} 
\not = \ebeta$.  Physically, this
corresponds to a situation of coexisting phases that normally
takes place at a first-order phase transition \cite{Touchette2004}. 

Before continuing with our analysis of ensemble equivalence, we 
make a number of basic definitions.  A function $f$ on $\R$ is said to be concave 
on $\R$ if $f$ maps $\R$ into $\R \cup \{-\infty\}$, $f \not \equiv 
-\infty$, and for all $u$ and $v$ in $\R$ and all $\lambda \in (0,1)$
\[
f(\lambda u + (1-\lambda)v) \geq \lambda f(u) + (1-\lambda) f(v) .
\]
Let $f \not \equiv -\infty$ be a function mapping $\R$ into
$\R \cup \{-\infty\}$.  We define $\mboxdomf$ to be the set of
$u$ for which $f(u) > -\infty$.  
For $\beta$ and $u$  in $\R$ the Legendre-Fenchel
transforms $f^*$ and $f^{**}$ are defined by 
\[
f^*(\beta) = \inf_{u \in \R} \{\lan \beta,u \ran - f(u)\}
\ \mbox{ and } \
f^{**}(u) = \inf_{\beta \in \R} \{\lan \beta,u \ran - f^*(\beta)\} .
\]
The function $f^*$ is concave and upper semicontinuous on $\R$
and for all $u$ we have $f^{**}(u) = f(u)$ if and only if 
$f$ is concave and upper semicontinuous on $\R$ \cite[Thm.\ VI.5.3]{Ell}.  When $f$ 
is not concave and upper semicontinuous, then 
$f^{**}$ is the smallest concave, upper semicontinuous function on $\R$
that satisfies $f^{**}(u) \geq f(u)$ for all $u$ \cite[Prop.\ A.2]{CosEllTouTur1}.
In particular, if for some $u$, $f(u) \not = f^{**}(u)$, then $f(u) < f^{**}(u)$. 

Let $f \not \equiv -\infty$ be a function mapping $\R$ into
$\R \cup \{-\infty\}$, $u$ a point in $\mboxdomf$,
and $K$ a convex subset of $\mboxdomf$.  We have the following
four additional definitions:
$f$ is concave at $u$ if $f(u) = f^{**}(u)$; $f$ 
is not concave at $u$ if $f(u) < f^{**}(u)$; 
$f$ is concave on $K$ if $f$ is concave at all $u \in K$;
and $f$ is strictly concave on $K$ if 
for all $u \not = v$ in $K$ and all $\lambda \in (0,1)$
\[
f(\lambda u + (1-\lambda)v) > \lambda f(u) + (1-\lambda) f(v) .
\]

We also introduce two sets
that play a central role in the theory.
Let $f$ be a concave function on $\R$
whose domain is an interval having nonempty interior.  
For $u \in \R$ the superdifferential of $f$ at $u$, denoted by $\partial f(u)$,
is defined to be the set of $\beta$ such that $\beta$ is the slope
of a supporting line of $f$ at $u$.
Any such $\beta$ is called a supergradient of $f$ at $u$.  
Thus, if $f$ is differentiable at $u \in \mboxintdomf$, 
then $\partial f(u)$ consists of the unique point $\beta = f'(u)$.
If $f$ is not differentiable at $u \in \mboxintdomf$, 
then $\mbox{dom}\, \partial f$
consists of all $\beta$ satisfying the inequalities 
\[
(f')^+(u) \leq \beta \leq (f')^-(u) ,
\]
where $(f')^-(u)$ and $(f')^+(u)$ denote the left-hand and right-hand
derivatives of $f$ at $u$.  The domain of $\partial f$, 
denoted by $\mbox{dom}\, \partial f$, is then defined to be the set of $u$ 
for which $\partial f(u) \not = \emptyset$.  

Complications arise because 
$\mbox{dom} \, \partial f$ can be a proper subset of $\mboxdomf$,
as simple examples clearly show.  Let $b$ be a
boundary point of $\mboxdomf$ for which $f(b) > -\infty$.  Then $b$ is in 
$\mbox{dom} \, \partial f$ if and only if the one-sided derivative of $f$ at $b$ is
finite.  For example, if $b$ is a left hand boundary point of $\mboxdomf$ and $(f')^+(b)$ is finite,
then $\partial f(b) = [(f')^+(b),\infty)$; any $\beta \in \partial f(b)$ is
the slope of a supporting line at $b$.
The possible discrepancy
between $\mbox{dom} \, \partial f$ and $\mboxdomf$ introduces unavoidable technicalities 
in the statements of several results concerning the existence of supporting lines.  

One of our goals is to find concavity and support
conditions on the microcanonical entropy guaranteeing that
the microcanonical and canonical 
ensembles are fully equivalent at all points $u \in \mbox{dom} \, s$ except possibly
boundary points.  If this is the case, then we say that the ensembles are
universally equivalent.  Here is a basic result in that direction. 
The universal equivalence stated in part (b) 
follows from part (a) and from part (a) of Theorem \ref{thm:equiv}.  The rest of the theorem
depends on facts concerning concave functions \cite[p.\ 1305]{CosEllTouTur1}.

\begin{thm}
\label{thm:strictlyconcave}
Assume that $\mboxemdoms$ is an interval having nonempty interior and that
$s$ is strictly concave on $\mboxemintdoms$ and continuous on $\mboxdoms$.
The following conclusions hold.

{\em (a)} $s$ has a strictly supporting line at all
$u \in \mboxemdoms$ except possibly boundary points.

{\em (b)} The microcanonical and canonical ensembles are universally 
equivalent; i.e., fully equivalent at all $u \in \mboxemdoms$ 
except possibly boundary points.  

{\em (c)} $s$ is concave on $\R$, and 
for each $u$ in part {\em (b)} the corresponding $\beta$ in the statement of 
full equivalence is any element of $\partial s(u)$.  

{\em (d)} If $s$ is differentiable
at some $u \in \mboxemdoms$, then the corresponding $\beta$ in part {\em (b)} is unique and is given by
the standard thermodynamic formula $\beta = s'(u)$.
\end{thm}

The next theorem extends Theorem
\ref{thm:equiv} by giving equivalence and
nonequivalence results involving $\eu$ and $\egbeta$, the sets of equilibrium
macrostates with respect to the microcanonical and Gaussian ensembles.  
The chief innovation is that $s(u)$ in Theorem \ref{thm:equiv} is replaced here by 
the generalized microcanonical entropy $s(u) - \gamma u^2$.
As we point out after the statement of Theorem \ref{thm:main}, for the purpose
of applications part (a) is its most important contribution.   
The usefulness of Theorem \ref{thm:main} is matched by the simplicity with 
which it follows from Theorem \ref{thm:equiv}.  Theorem \ref{thm:main}
is a special case of Theorem 3.4 in \cite{CosEllTouTur1}, obtained
by specializing the generalized canonical ensemble and the associated
set of equilibrium macrostates to the Gaussian ensemble and the
set $\egbeta$ of Gaussian equilibrium macrostates.

\begin{theorem}
\label{thm:main} 
Given $\gamma \geq 0$, define $s_\gamma(u) = s(u) - \gamma u^2$.
In parts {\em (a)}, {\em (b)}, and {\em (c)}, $u$ denotes
any point in $\mbox{\em{dom}} \, s$.

{\em (a)}  \mbox{{\em {\bf Full equivalence.}}} \ There exists $\beta$
such that ${\cal E}^u = {\cal E}_{\beta,\gamma}$ 
if and only if $s_\gamma$ has a strictly supporting line at $u$ with slope $\beta$. 

{\em (b)} \mbox{{\em {\bf Partial equivalence.}}} \ There exists $\beta$ such that
${\cal E}^u \subset {\cal E}_{\beta,\gamma}$ but ${\cal E}^u \not = {\cal E}_{\beta,\gamma}$
if and only if $s_\gamma$ has a nonstrictly supporting line at $u$ with slope $\beta$. 

{\em (c)} \mbox{{\em {\bf Nonequivalence.}}} \ For all $\beta$,
${\cal E}^u \cap {\cal E}_{\beta,\gamma} = \emptyset$
if and only if $s_\gamma$ has no supporting line at $u$.
\iffalse
Except possibly for relative boundary points of $\mboxemdoms$,
the latter condition is equivalent to the nonconcavity of $s_\gamma$ at $u$.  
\fi

{\em (d)} \mbox{{\em {\bf Gaussian is always realized microcanonically.}}}
\ For any  $\beta$ we have $\tilde{H}({{\cal E}}_{\beta,\gamma})
\subset \mbox{{\em dom}} \, s$ and 
\[
{\cal E}_{\beta,\gamma} = \bigcup_{u \in \thi({{\cal E}}_{\beta,\gamma})} 
{\cal E}^u .
\]
\end{theorem}

\noi
{\bf Proof.}  For $\gamma \geq 0$ and $B \in {\cal F}_n$ we define a new probability measure
\[
P_{n,\gamma}\{B\} = \frac{1}{\displaystyle
\int_{\Omega_n} \exp[-a_n \gamma h_n^2] \, dP_n} \cdot 
\int_B \exp[-a_n \gamma h_n^2] \, dP_n .
\]
With respect to $P_{n,\gamma}$, $Y_n$ 
satisfies the LDP on $\X$ with rate function
\[
I_\gamma(x) = I(x) + \gamma [\thi(x)]^2 - \psi(\gamma),
\]
where $\psi(\gamma) = \inf_{y \in \X}\{I(y) + \gamma [\thi(y)]^2\}$.
Replacing the prior measure $P_n$
in the  canonical ensemble with $P_{n,\gamma}$ gives the Gaussian
ensemble $P_{n,\beta,\gamma}$, which has $\egbeta$ as the associated
set of equilibrium macrostates. 
On the other hand, replacing the prior measure $P_n$ 
in the  microcanonical ensemble with $P_{n,\gamma}$ gives
\[
P_{n,\gamma}^{u,r}\{B\} = P_{n,\gamma}\{B \mid h_n \in [u-r,u+r]\} ,
\]
By continuity, for $\omega$ satisfying $h_n(\omega) \in [u-r,u+r]$,
$[h_n(\omega)]^2$ converges to $u^2$ uniformly in $\omega$ and $n$ as $r \goto 0$.  
It follows that with respect to $P_{n,\gamma}^{u,r}$, $Y_n$ satisfies the LDP on $\X$, in 
the double limit $n \goto \infty$ and $r \goto 0$, with the same rate function
$I^u$ as in the LDP for $Y_n$ with respect to $P_n^{u,r}$.  As a result,
the set of equilibrium macrostates corresponding to 
$P_{n,\gamma}^{u,r}$ coincides with the set $\eu$ of microcanonical equilibrium macrostates.  

It follows from parts (a)--(c) of
Theorem \ref{thm:equiv} that all equivalence and nonequivalence relationships
between $\eu$ and $\egbeta$ are expressed
in terms of support properties of the function $\tilde{s}_\gamma$ obtained
from $s$ by replacing the rate function $I$ 
by the new rate function $I_\gamma$.  The function $\tilde{s}_\gamma$ is given by
\beas
\tilde{s}_\gamma(u) & = & - \inf\{I_\gamma(x) : x \in \X, \tilde{H}(x) = u\} \\
& = & -\inf\{I(x) + \gamma \tilde{H}(x)^2 : x \in \X, \tilde{H}(x) = u\} + \psi(\gamma) \\
& = & s(u) - \gamma u^2 + \psi(\gamma) .
\eeas
Since $\tilde{s}_\gamma(u)$ differs from $s_\gamma(u) =
s(u)  - \gamma u^2$ by the constant $\psi(\gamma)$, we conclude that all equivalence and nonequivalence
relationships between $\mathcal{E}^u$ and $\egbeta$ are expressed
in terms of the same support properties of $s_\gamma$. 
This completes the derivation of parts (a)--(c) of
Theorem \ref{thm:main} from parts (a)--(c) of Theorem \ref{thm:equiv}. 
Similarly, part (d) of Theorem \ref{thm:main} follows from part (d) of 
Theorem \ref{thm:equiv}. \ \ink

\skp

The importance of part (a) of Theorem \ref{thm:main} in applications
is emphasized by the following theorem, which will be applied in the sequel.  
This theorem is the analogue of Theorem \ref{thm:strictlyconcave} for
the Gaussian ensemble, $s$ in that theorem being replaced by $s_\gamma$.  
The functions $s$ and $s_\gamma$ have the same domains. 
The universal equivalence stated in part (b) of the next theorem follows
from part (a) and from part (a) of Theorem \ref{thm:main}.

\begin{thm}
\label{thm:sgstrictlyconcave}
For $\gamma \geq 0$, define $s_\gamma(u) = s(u) - \gamma u^2$.
Assume that $\mboxemdoms$ is an interval having nonempty interior and that
$s_\gamma$ is strictly concave on $\mboxemintdoms$ and continuous on $\mboxdoms$.
The following conclusions hold.

{\em (a)} $s_\gamma$ has a strictly supporting line at all
$u \in \mboxemdoms$ except possibly boundary points.

{\em (b)} The microcanonical ensemble and the Gaussian ensemble
defined in terms of this $\gamma$ are universally 
equivalent; i.e., fully equivalent at all $u \in \mboxemdoms$ 
except possibly boundary points.  

{\em (c)} $s_\gamma$ is concave on $\R$, and
for each $u$ in part {\em (b)} the corresponding $\beta$ in the statement of 
full equivalence is any element of $\partial s_\gamma(u)$.  

{\em (d)} If $s_\gamma$ is differentiable
at some $u \in \mboxemdoms$, then the corresponding $\beta$ in 
part {\em (b)} is unique and is given by
the thermodynamic formula $\beta = s_\gamma'(u)$.
\end{thm}

The most important repercussion of Theorem \ref{thm:sgstrictlyconcave} is the ease
with which one can prove that the microcanonical and Gaussian 
ensembles are universally equivalent in those cases 
in which the microcanonical and canonical
ensembles are not fully or partially equivalent.  
This rests mainly on part (b) of Theorem \ref{thm:sgstrictlyconcave}, which
states that universal equivalence of ensembles holds if 
there exists a $\gamma \geq 0$ such that $s_\gamma$ is strictly concave on $\mboxintdoms$.
The existence of such a $\gamma$ follows from a natural set of hypotheses on $s$
stated in Theorem \ref{thm:applysigma1} in the next section.

\section{Universal Equivalence via the Generalized Canonical Ensemble}
\setcounter{equation}{0}
\label{section:gencanon}

This section addresses a basic foundational issue in statistical mechanics.  
Under the assumption that the microcanonical entropy 
is $C^2$ and $s''$ is bounded above, we show in Theorem \ref{thm:applysigma1}
that when the canonical ensemble is nonequivalent to the microcanonical ensemble
on a subset of energy values $u$, 
it can often be replaced by a Gaussian ensemble that is 
univerally equivalent to the microcanonical ensemble;
i.e., fully equivalent at all $u \in \mboxdoms$ except possibly boundary points.
Theorem \ref{thm:localapply} is a weaker version that can often be 
applied when $s''$ is not bounded above.
In the last section of the paper, these results will be illustrated in the context of
the Curie-Weiss-Potts lattice-spin model.

In Theorem \ref{thm:applysigma1} the strategy is to find a quadratic function $\gamma u^2$ such
that $s_\gamma(u) = s(u) - \gamma u^2$ is strictly concave on $\mboxintdoms$ and continuous on $\mboxdoms$.
Parts (a) and (b) of Theorem \ref{thm:sgstrictlyconcave} then yields the universal equivalence.  As the next 
proposition shows, an advantage of working with quadratic functions is that
support properties of $s_\gamma$ involving a supporting line are equivalent
to support properties of $s$ involving a supporting parabola defined in
terms of $\gamma$.  This observation gives a geometrically intuitive way
to find a quadratic function guaranteeing universal ensemble equivalence.

In order to state the proposition, we need a definition.
Let $f$ be a function mapping $\R$ into $\R \cup \{-\infty\}$,
$u$ and $\beta$ points in $\R$, and $\gamma \geq 0$.  
We say that $f$ has a supporting
parabola at $u$ with parameters $(\beta,\gamma)$ if 
\be
\label{eqn:parabola}
f(v) \leq  f(u) + \langle \beta, v-u \rangle + \gamma(v-u)^2 \ \mbox{ for all } v .
\ee
The parabola is said to be strictly supporting
if the inequality is strict for all $v \not = u$.

\begin{prop}
\label{prop:parabola}
$f$ has a {\em (}strictly{\em )} 
supporting parabola at $u$ with parameters $(\beta,\gamma)$ if and
only if $f - \gamma (\cdot)^2$ has a {\em (}strictly{\em )} supporting line at $u$ with slope
$\tilde\beta$.  The quantities 
$\beta$ and $\tilde{\beta}$ are related by $\tilde{\beta} = \beta
- 2\gamma u$.
\end{prop}

\noi
{\bf Proof.}  The proof is based on the identity 
$(v-u)^2 = v^2 - 2u(v - u) - u^2$.
If $f$ has a strictly supporting
parabola at $u$ with parameters $(\beta,\gamma)$,
then for all $v \not = u$  
\[
f(v) - \gamma v^2 < f(u) - \gamma u^2 + \tilde{\beta}(v - u) ,
\]
where $\tilde{\beta} = \beta - 2\gamma u$.  Thus $f - \gamma (\cdot)^2$ has a 
strictly supporting
line at $u$ with slope $\tilde{\beta}$.  The converse is proved 
similarly, as is the case in which the supporting line or parabola
is supporting but not strictly supporting.  \ \ink

\skp
The first application of Theorem \ref{thm:sgstrictlyconcave}
is Theorem \ref{thm:applysigma1}, which
gives a criterion guaranteeing the existence of a
quadratic function $\gamma u^2$ such that $s_\gamma(u) = s(u) - \gamma u^2$ is
strictly concave on $\mboxdoms$.
The criterion --- that 
$s''$ is bounded above on the interior of $\mboxdoms$ --- is essentially
optimal for the existence of a fixed
quadratic function $\gamma u^2$ guaranteeing the strict concavity of $s_\gamma$. 
The situation
in which $s''$ is not bounded above on the interior of $\mboxdoms$
can often be handled by Theorem \ref{thm:localapply}, which is a local version of
Theorem \ref{thm:applysigma1}.   

\begin{thm}
\label{thm:applysigma1}
Assume that $\mboxdoms$ is an interval having nonempty interior.
Assume also that $s$ is continuous on $\mboxemdoms$, $s$ is twice continuously differentiable
on $\mboxemintdoms$, and $s''$ 
is bounded above on $\mboxemintdoms$.  
Then for all sufficiently large $\gamma \geq 0$, 
conclusions {\em (a)--(c)} hold.  Specifically, if $s$ is strictly concave
on $\mboxemdoms$, then we choose any $\gamma \geq 0$, and otherwise we choose
\be
\label{eqn:gamma0sigma1}
\gamma > \gamma_0 = \textstyle \frac{1}{2} \cdot 
\displaystyle \sup_{u \in \mbox{\scriptsize {\em int(dom}} \, s)} s''(u) .
\ee

{\em (a)}  $s_\gamma(u) = s(u) - \gamma u^2$ is strictly concave and continuous on $\mboxemdoms$.

{\em (b)} $s_\gamma$ has a 
strictly supporting line, and $s$ has a strictly supporting 
parabola, at all $u \in \mboxemdoms$ except possibly
boundary points.  At a boundary point $s_\gamma$ has a strictly supporting line, and $s$
has a strictly supporting parabola, 
if and only if the one-sided derivative of $s_\gamma$
is finite at that boundary point.

{\em (c)} The microcanonical ensemble and the Gaussian
ensemble defined in terms of this $\gamma$ are universally equivalent; 
i.e., fully equivalent at all $u \in \mboxemdoms$ except possibly boundary points.
For all $u \in \mboxemintdoms$ the value of $\beta$ defining the universally
equivalent Gaussian ensemble is 
unique and is given by $\beta = s'(u) - 2\gamma u$. 

\end{thm}

\noi
{\bf Proof.}   
(a) If $s$ is strictly concave on $\mboxdoms$, then 
$s_\gamma$ is also strictly concave on this set for 
any $\gamma \geq 0$.  We now consider the case in which $s$ is not strictly concave 
on $\mboxdoms$.  For any $\gamma \geq 0$, $s_\gamma$ is continuous on $\mboxdoms$.  
If, in addition, we choose $\gamma > \gamma_0$ in accordance with (\ref{eqn:gamma0sigma1}),
then for all $u \in \mboxintdoms$
\[
s_\gamma''(u) = s''(u) - 2 \gamma < 0 .
\]
A straightforward extension of the proof of 
Theorem 4.4 in \cite{Rock}, in which the inequalities
in the first two displays are replaced by strict inequalities,
shows that $-s_\gamma$ is strictly convex on $\mboxintdoms$ and thus
that $s_\gamma$ is strictly concave on $\mboxintdoms$. 
If $s_\gamma$ is not strictly concave on $\mboxdoms$, 
then $s_\gamma$ must be affine on an interval.  Since this
violates the strict concavity on $\mboxintdoms$, part (a) is proved.

(b) The first assertion follows from part (a) of the present theorem,
part (a) of Theorem \ref{thm:sgstrictlyconcave}, and Proposition \ref{prop:parabola}.
Concerning the second assertion about boundary points, the reader is
referred to the discussion before Theorem \ref{thm:strictlyconcave}.

(c) The universal equivalence of the two ensembles is a consequence
of part (a) of the present theorem and part (b) of Theorem \ref{thm:sgstrictlyconcave}.
The full equivalence of the ensembles at all $u \in \mboxintdoms$
is equivalent to the existence of a strictly
supporting line at each $u \in \mboxintdoms$ [Thm.\ \ref{thm:main}(a)].  
Since $s_\gamma(u)$ is differentiable at all $u \in \mboxintdoms$, for each
$u$ the slope of the strictly supporting line at $u$ is unique
and equals $s_\gamma'(u)$
\cite[Thm. A.1(b)]{CosEllTouTur1}. \ \ink

\skp

Suppose that $s$ is $C^2$ on the interior of $\mboxdoms$ but
the second-order partial derivatives of $s$ are not bounded above.
This arises, for example, in the Curie-Weiss-Potts model, in which
$\mbox{dom} \, s$ is a closed, bounded interval of $\R$ and 
$s''(u) \goto \infty$ as $u$ approaches the right hand endpoint of
$\mbox{dom} \, s$ [see \S \ref{section:cwp}].
In such cases one cannot expect that the conclusions of Theorems
\ref{thm:applysigma1} will be satisfied; 
in particular, that there exists $\gamma \geq 0$ 
such that $s_\gamma(u) = s(u) - \gamma u^2$ has a strictly
supporting line at each point of the interior of $\mboxdoms$ and thus that
the ensembles are universally equivalent.

In order to overcome this difficulty, 
we introduce Theorem \ref{thm:localapply}, a local version of Theorem \ref{thm:applysigma1}.  
Theorem \ref{thm:localapply} handles the case in which $s$ is $C^2$
on an open set $K$ but either $K$ is not all of $\mboxintdoms$
or $K = \mboxintdoms$ and the second-order partial
derivatives of $s$ are not all bounded above on $K$.
In neither of these situations are the hypotheses of Theorem \ref{thm:applysigma1} satisfied.  

In Theorem \ref{thm:localapply} other hypotheses are given 
guaranteeing that for each $u \in K$ 
there exists $\gamma$  such that
$s_\gamma$ has a strictly supporting line at $u$; in general, $\gamma$
depends on $u$.
However, with the same $\gamma$, $s_\gamma$ might also have a strictly supporting
line at other values of $u$.  In general, as one increases $\gamma$, 
the set of $u$ at which $s_\gamma$ has a strictly supporting line cannot decrease.
Because of part (a) of Theorem \ref{thm:main}, this can be restated in terms of
ensemble equivalence involving the set $\egammabeta$ of Gaussian
equilibrium macrostates.  Defining
\[
F_\gamma = \{u \in K : \mbox{there exists } \beta \mbox{ such that } \egammabeta = \eu\} ,
\]
we have $F_{\gamma_1} \subset F_{\gamma_2}$ whenever $\gamma_2 >
\gamma_1$ and because of Theorem \ref{thm:localapply},
$\bigcup_{\gamma > 0} F_\gamma = K$. 
This phenomenon is investigated in Section \ref{section:cwp} for the Curie-Weiss-Potts model.

In order to state Theorem \ref{thm:localapply}, we define
for $u \in K$ and $\lambda \geq 0$ 
\[
\label{dx0s}
D(u,s'(u), \lambda) = \left\{v \in \mboxdoms : s(v) \geq s(u) + 
s'(u)(v - u) + \lambda (v - u)^2 \right\} .
\]
Geometrically, this set contains all points for which the 
parabola with parameters $(s'(u),\lambda)$ passing through $(u,s(u))$
lies below the graph of $s$.
Clearly, since $\lambda \geq 0$, we have
$D(u,s'(u), \lambda)$ $\subset D(u,s'(u),0)$;
the set $D(u,s'(u),0)$ contains
all points for which the 
graph of the line with slope $s'(u)$ 
passing through $(u,s(u))$ lies below the 
graph of $s$.  Thus, in the next theorem
the hypothesis that for each $u \in K$ the set 
$D(u,s'(u),\lambda)$ is bounded for some $\lambda \geq 0$
is satisfied if $\mboxdoms$ is bounded or, more generally, if
$D(u,s'(u),0)$ is bounded.  
The latter set is bounded if, for example, $-s$ is superlinear; 
i.e., \[
\lim_{|v| \goto \infty} s(v)/|v| = -\infty .
\]
The quantity $\gamma_0(u)$ appearing in the next theorem is defined in
equation (5.7) in \cite{CosEllTouTur1}.

\begin{thm}
\label{thm:localapply}
Let $K$ an open subset of $\mbox{{\em dom}} \, s$ and 
assume that $s$ is twice continuously differentiable on $K$.
Assume also that $\mboxemdoms$ is bounded or, more generally, that
for every $u \in \mbox{{\em int}} \, K$ there exists $\lambda \geq 0$ such that 
$D(u,s'(u), \lambda)$ is bounded.
Then for each $u \in K$ there exists 
$\gamma_0(u) \geq 0$ with the following properties.

{\em (a)} For each $u \in K$ and any $\gamma > \gamma_0(u)$, $s$ has a strictly supporting
parabola at $u$ with parameters $(s'(u),\gamma)$. 

{\em (b)} For each $u \in K$ and any 
$\gamma > \gamma_0(u)$,
$s_\gamma = s - \gamma (\cdot)^2$ has a strictly supporting line at $u$ with 
slope $s'(u) - 2 \gamma u$.

{\em (c)} For each $u \in K$  and any $\gamma >
\gamma_0(u)$,
the microcanonical ensemble and the Gaussian
ensemble defined in terms of this $\gamma$ are fully equivalent at $u$.  
The value of $\beta$ defining the Gaussian ensemble is unique and is given by
$\beta = s'(u) - 2\gamma u$.  
\end{thm}

\noi
{\bf Comments on the Proof.} (a)  
We first choose a parabola that is strictly supporting in a neighborhood of $u$
and then adjust $\gamma$ so that the parabola becomes
strictly supporting on all $\R$.  Proposition \ref{prop:parabola}
guarantees that $s - \gamma (\cdot)^2$ has a strictly
supporting line at $u$.   Details are given in \cite[pp.\ 1319--1321]{CosEllTouTur1}.

(b) This follows from part (a) of the present theorem and 
Proposition \ref{prop:parabola}.

(c) For $u \in K$ the full equivalence of the ensembles follows from part (b) of the present
theorem and part (a) of Theorem
\ref{thm:main}.  The value of $\beta$ defining the fully equivalent
Gaussian ensemble is determined by a routine argument given in
\cite[p.\ 1321]{CosEllTouTur1}.  \ \ink

\skp

Theorem \ref{thm:localapply} suggests an extended form of the notion of universal
equivalence of ensembles.  In Theorem \ref{thm:applysigma1} we 
are able to achieve full equivalence of ensembles for all 
$u \in \mboxdoms$ except possibly
boundary points by choosing an appropriate $\gamma$ that
is valid for all $u$.  This leads to the observation that
the microcanonical ensemble and the Gaussian 
ensemble defined in terms of this $\gamma$ are universally
equivalent.  In Theorem \ref{thm:localapply}
we can also achieve full equivalence of ensembles
for all $u \in K$.  However, in contrast to 
Theorem \ref{thm:applysigma1},
the choice of $\gamma$ for which the two ensembles are
fully equivalent depends on $u$.  We summarize the ensemble equivalence
property articulated in part (c) of Theorem \ref{thm:localapply} by 
saying that relative to the set of quadratic functions,
the microcanonical and Gaussian
ensembles are universally equivalent on the open set $K$ of energy values.

We complete our discussion of the generalized canonical ensemble and its
equivalence with the microcanonical ensemble by noting that 
the smoothness hypothesis on $s$ in Theorem \ref{thm:localapply}
is essentially satisfied whenever
the microcanonical ensemble exhibits no phase transition at any $u \in K$.
In order to see this, 
we recall that a point $u_c$ at which $s$ is not differentiable represents
a first-order, microcanonical phase transition \cite[Fig.\ 3]{ETT}.  
In addition, a point $u_c$ at which $s$ is differentiable
but not twice differentiable represents a second-order, microcanonical phase transition
\cite[Fig.\ 4]{ETT}.  It follows that $s$ is smooth on any open set $K$ not containing 
such phase-transition points.  Hence, if
the other conditions in Theorem \ref{thm:localapply} are valid, then 
the microcanonical and Gaussian ensembles are universally equivalent on $K$ relative
to the set of quadratic functions.  In particular, if the microcanonical ensemble
exhibits no phase transitions, then $s$ is smooth on all of $\mboxintdoms$.  This implies the 
universal equivalence of the two ensembles provided that the other
conditions are valid in Theorem \ref{thm:applysigma1}.

In the next section we apply the results in this
paper to the Curie-Weiss-Potts model. 

\section{Applications to the Curie-Weiss-Potts Model}
\setcounter{equation}{0}
\label{section:cwp}

The Curie-Weiss-Potts model is a mean-field approximation to the nearest-neighbor
Potts model, which takes its place next to the Ising model as one of the most
versatile models in equilibrium statistical mechanics \cite{Wu}.  Although the Curie-Weiss-Potts
model is considerably simpler to analyze, it is an excellent model to illustrate the general
theory presented in this paper, lying at the boundary of the set of models for which a 
complete analysis involving explicit formulas is available.  As we will see, there
exists an interval $N$ such that for any $u \in N$ the microcanonical ensembe is
nonequivalent to the canonical ensemble. 
The main result, stated in Theorem \ref{thm:fgamma}, is that for any $u \in N$
there exists $\gamma \geq 0$ such that the microcanonical ensemble
and the Gaussian ensemble defined in terms of this $\gamma$
are fully equivalent for all $v \leq u$.  While not as strong as universal equivalence,
the ensemble equivalence proved in Theorem \ref{thm:fgamma}
is considerably stronger than the local equivalence
stated in Theorem \ref{thm:localapply}.

Let $q \geq 3$ be a fixed integer and define 
$\Lambda = \{\theta^1,\theta^2,\ldots,\theta^q\}$, where the 
$\theta^i$ are any $q$ distinct vectors in $\R^q$.  
In the definition of the Curie-Weiss-Potts model, 
the precise values of these vectors is immaterial.
For each $n \in \N$ the model is defined by spin random variables
$\omega_1,\omega_2,\ldots,\omega_n$ that take values in $\Lambda$.
The ensembles for the model are defined in terms of
probability measures on the configuration spaces $\Lambda^n$, which consist of 
the microstates $\omega = (\omega_1,\omega_2,\ldots,\omega_n)$.   
We also introduce the $n$-fold product measure $P_n$ 
on $\Lambda^n$ with identical one-dimensional marginals 
\[
\bar\rho = \frac{1}{q}\sum_{i=1}^q \delta_{\theta^i} .
\]
Thus for all $\omega \in \Lambda^n$, $P_n(\omega) = \frac{1}{q^n}$.
For $n \in \N$ and $\omega \in \Lambda^n$ 
 the Hamiltonian for the $q$-state Curie-Weiss-Potts model 
is defined by 
\[
H_n(\omega) = - \frac{1}{2n} \sum_{j,k=1}^n \delta(\omega_j,\omega_k) ,
\]
where $\delta(\omega_j,\omega_k)$ equals 1 if $\omega_j = \omega_k$ and
equals 0 otherwise.  The energy per particle is defined by
$h_n(\omega) = \frac{1}{n} H_n(\omega)$.

With this choice of $h_n$ and with $a_n = n$, 
the microcanonical, canonical, and Gaussian ensembles for the model 
are the probability measures on $\Lambda^n$ defined as in 
(\ref{eqn:microens}), (\ref{eqn:canonens}), and
(\ref{eqn:gencanon}).  The key to our analysis of the Curie-Weiss-Potts model is to express 
$h_n$ in terms of the macroscopic variables 
\[
L_n = L_n(\omega) = (L_{n,1}(\omega),L_{n,2}(\omega),\ldots,L_{n,q}(\omega)) ,
\]
the $i$th component of which is defined by
\[
\label{eqn:empmeasure}
L_{n,i}(\omega) = \frac{1}{n} \sum_{j=1}^n
\delta(\omega_j,\theta^i) .
\]
This quantity equals the relative frequency with which 
$\omega_j, j \in \{1,\ldots,n\},$ equals $\theta^i$.
The empirical vectors $L_n$ take values in the set of probability vectors
\[
\mathcal{P} = \left\{\nu \in \R^q : \nu = (\nu_1,\nu_2,\ldots,\nu_q), \mbox{ each } \nu_i \geq 0,
\sum_{i=1}^q \nu_i = 1 \right\}.
\]
Each probability vector in $\cp$ represents a possible equilibrium
macrostate for the model.  

There is a one-to-one correspondence between $\cp$ and the set $\cp(\Lambda)$
of probability measures on $\Lambda$, $\nu \in \cp$ corresponding to the 
probability measure $\sum_{i=1}^q \nu_i \delta_{\theta^i}$.  The element
$\rho \in \cp$ corresponding to the one-dimensional marginal $\bar\rho$ of
the prior measures $P_n$ is the uniform vector having equal components $\frac{1}{q}$.
For $\omega \in \Lambda^n$ the element of $\cp$ corresponding to the 
empirical vector $L_n(\omega)$ is the empirical measure of
the spin random variables $\omega_1,\omega_2,\ldots,\omega_n$.

We denote by $\lan \cdot, \cdot \ran$ the inner product on $\R^q$.  Since
\[
\sum_{i=1}^q \sum_{j=1}^n \delta(\omega_j,\xi^i) \cdot \sum_{k=1}^n \delta(\omega_k,\xi^i) =
\sum_{j,k=1}^n \delta(\omega_j,\omega_k) ,
\]
it follows that the energy per particle can be rewritten as 
\[
\label{eqn:empiricalmeasureenergy}
h_{n}(\omega) = - \frac{1}{2n^2} \sum_{j,k=1}^n \delta(\omega_j,\omega_k) =
-\ts\frac{1}{2} \lan L_n(\omega),L_n(\omega) \ran ,
\]
i.e., 
\[
\label{eqn:interactfcn}
h_n(\omega) = \thi(L_n(\omega)), \mbox{ where } 
\thi(\nu) = -\ts\frac{1}{2} \lan \nu,\nu \ran  
\mbox{ for } \nu \in \mathcal{P} .
\]
$\thi$ is the energy representation function for the model.

In order to define the sets
of equilibrium macrostates with respect to the three ensembles, 
we appeal to Sanov's Theorem.  This states that with respect to the
product measures $P_n$, the empirical vectors
$L_n$ satisfy the LDP on $\cp$ with rate function
given by the relative entropy $R(\cdot|\rho)$ \cite[Thm.\ VIII.2.1]{Ell}.  
For $\nu \in \cp$ this is defined by 
\[
R(\nu|\rho) = \sum_{i=1}^q \nu_i \log (q\nu_i) .
\]
With the choices $I = R(\cdot|\rho)$, $\thi = -\frac{1}{2}\lan \cdot,\cdot \ran$, 
and $a_n = n$,
$L_n$ satisfies the LDP on $\cp$ with respect to each of the three
ensembles with the rate functions given by (\ref{eqn:iu}), (\ref{eqn:ibeta}),
and (\ref{eqn:ibetag}).  In turn, the corresponding
sets of equilibrium macrostates are given by
\[
\label{eqn:micromacro}
\eu = 
\left\{ \nu \in \mathcal{P}
: R(\nu | \rho) \mbox{ is minimized subject to } 
\ts \thi(\nu) = u \! \right\} .
\]
\[
\label{eqn:canonmacro}
\ebeta =  \left\{\nu \in
\mathcal{P} : R(\nu |\rho) + \ts \beta \thi(\nu) \mbox{ is minimized } \! \right\},
\]
and
\[
\label{eqn:gaussianmacro}
\egbeta =  \left\{\nu \in
\mathcal{P} : R(\nu |\rho) + \ts \beta\thi(\nu) 
+ \gamma [\thi(\nu)]^2 \mbox{ is minimized } \! \right\} ,
\]
Each element $\nu$ in $\eu$, $\ebeta$, and $\egbeta$ 
describes an equilibrium configuration of the model with respect to the corresponding
ensemble in the thermodynamic limit.  
The $i$th component $\nu_i$ gives the asymptotic relative
frequency of spins taking the value $\theta^i$.  

\iffalse
The set $\eu$ is defined for all $u$ for which the constraint in the definition
of $I^u$ is satisfied for some $\nu \in \cp$.  Otherwise, $\eu$ is not defined.
If $\eu$ is defined, then $\eu$ is nonempty; if $\eu$ is not defined, 
then we set $\eu = \emptyset$. 
\fi

As in (\ref{eqn:entropy}), the microcanonical entropy is defined by
\[
\label{eqn:cwpentropy}
s(u) = - \inf\{R(\nu|\rho) \, : \, \nu \in \cp, \thi(\nu) = u\} .
\]
Since $R(\nu|\rho) < \infty$ for all $\nu \in \cp$, $\mboxdoms$ equals
the range of $\thi(\nu) = -\frac{1}{2} \lan \nu,\nu \ran$ on $\cp$,
which is the closed interval $[-\frac{1}{2},-\frac{1}{2q}]$.
The set $\eu$ of microcanonical equilibrium macrostates is nonempty
precisely for $u \in \mboxdoms$.  
For $q = 3$, the microcanonical entropy can be determined explicitly. 
For all $q \geq 4$ the microcanonical entropy can also be determined explicitly
provided Conjecture 4.1 in \cite{CosEllTou1} is valid; this conjecture has been
verified numerically for all $q \in \{4,5,\ldots,10^4\}$.  The formulas 
for the microcanonical entropy are given in Theorem 4.3 in \cite{CosEllTou1}.

We first consider the relationships between $\eu$ and $\ebeta$, which according to Theorem
\ref{thm:equiv} are determined by 
support properties of $s$. These properties can be seen in Figure 1.  The quantity
$u_0$ appearing in this figure equals $[-q^2 + 3q - 3]/[2q(q-1)]$ \cite[Lem.\ 6.1]{CosEllTou1}.
Figure 1 is not the actual graph of $s$ but a schematic graph that accentuates the shape of 
the graph of $s$ together with
the intervals of strict concavity and nonconcavity of this function.  

These and other details of the graph of $s$ 
are also crucial in analyzing the relationships between $\eu$ and $\egbeta$. 
Denote $\mboxdoms$ by $[u_\ell,u_r]$, where $u_\ell = -\frac{1}{2}$ and $u_r = -\frac{1}{2q}$.
These details include the observation that there exists $w_0 \in (u_0,u_r)$ such 
that $s$ is a concave-convex function with break point $w_0$; i.e.,
the restriction of $s$ to $(u_\ell,w_0)$ is strictly concave 
and the restriction of $s$ to $(w_0,u_r)$ is strictly convex.  
A difficulty with this determination is that for certain values of $q$, including
$q = 3$, the intervals of strict concavity and strict convexity
are shallow and therefore difficult to discern.  Furthermore, 
what seem to be strictly concave and strictly convex portions of 
this function on the scale of the entire graph might reveal themselves
to be much less regular on a finer scale.  Conjecture \ref{conj:sprimeprimeprime}
gives a set of properties of $s$ implying there exists $w_0 \in (u_0,u_r)$ such 
that $s$ is a concave-convex function with break point $w_0$.  In particular, this property
of $s$ guarantees that $s$ has the support properties stated in the three items appearing
in the next paragraph. Conjecture
\ref{conj:sprimeprimeprime} has been verified numerically for all $q \in \{4,5,\ldots,10^4\}$.

\begin{figure}[t]
\label{figure:entropy}
\begin{center}
\includegraphics{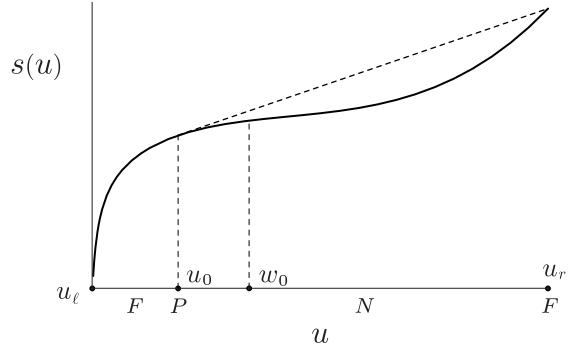}
\caption{ \small  Schematic graph of $s(u)$, showing the set
$F = (u_\ell, u_0) \cup \{u_r\}$ of full ensemble
equivalence, the singleton set $P = \{u_0\}$ of partial equivalence, and the set
$N = (u_0,u_r)$ of nonequivalence, where $u_\ell = -\frac{1}{2}$ and $u_r = -\frac{1}{2q}$.
For $u \in F \cup P = (u_\ell, u_0] \cup \{u_r\}$,
$s(u) = s^{**}(u)$; for $u \in N$, $s(u) < s^{**}(u)$ and the graph of $s^{**}$
consists of the dotted line segment with slope $\beta_c$.  The slope of $s$ 
at $u_\ell$ is $\infty$.  The quantity $w_0$ is discussed after
Conjecture \ref{conj:sprimeprimeprime}.}  
\end{center}
\end{figure} 

We define the sets 
\[
F = (u_\ell, u_0) \cup \{u_r\}, \ P = \{u_0\}, \ and \ 
N = (u_0,u_r) .
\]
Figure 1 and Theorem \ref{thm:equiv} then show that these sets are 
respectively the sets of full equivalence, partial equivalence,
and nonequivalence of the microcanonical and
canonical ensembles.  The details are given in the next three items.
In Theorem 6.2 in \cite{CosEllTou1} all these conclusions concerning ensemble
equivalence and nonequivalence are proved analytically without reference to 
the form of $s$ given in Figure 1. 

\begin{enumerate}
\item $s$ is strictly concave on the interval $(u_\ell, u_0)$ and 
has a strictly supporting line at each $u \in (u_\ell, u_0)$ and at $u_r$.  Hence
for $u \in F = (u_\ell, u_0) \cup \{u_r\}$ 
the ensembles are fully equivalent in the sense that there exists
$\beta$ such that $\eu = \ebeta$ [Thm.\ \ref{thm:equiv}(a)]. 
\item $s$ is concave but not strictly concave
at $u_0$ and has a nonstrictly supporting line at $u_0$ that also touches the graph
of $s$ over the right hand endpoint $u_r$.  Hence for $u \in P = \{u_0\}$ the
ensembles are partially equivalent in the sense that there exists
$\beta$ such that $\eu \subset \ebeta$ but $\eu \not = \ebeta$
[Thm.\ \ref{thm:equiv}(b)].  
\item $s$ is not concave on $N = (u_0,u_r)$ and has no supporting line 
at any $u \in N$.  Hence
for $u \in N$ the ensembles are nonequivalent in the sense that 
for all $\beta$, $\eu \cap \ebeta = \emptyset$ [Thm.\ \ref{thm:equiv}(c)].
\end{enumerate}

The explicit calculation of the elements of $\ebeta$ and $\eu$ given in \cite{CosEllTou1}
shows different continuity properties of these two sets.
$\ebeta$ undergoes a discontinuous phase transition
as $\beta$ increases through the critical inverse temperature
$\beta_c = \frac{2(q-1)}{q-2}\log(q-1)$, the unique
macrostate $\rho$ for $\beta < \beta_c$ bifurcating discontinuously into the $q$ distinct macrostates
for $\beta > \beta_c$.
By contrast, $\eu$ 
undergoes a continuous phase transition as $u$ decreases from the maximum
value $u_r = -\frac{1}{2q}$, the unique
macrostate $\rho$ for $u = u_r$ bifurcating continuously into the $q$ distinct macrostates for 
$u < u_r$.
The different continuity properties of these phase transitions shows already that the
canonical and microcanonical ensembles are nonequivalent. 

For $u$ in the interval $N$ of ensemble nonequivalence, 
the graph of $s^{**}$ is affine; this is depicted by the 
dotted line segment in Figure 1.  One can show that the slope of the affine portion
of the graph of $s^{**}$ equals the critical inverse temperature $\beta_c$.  

This completes the discussion of the equivalence and nonequivalence of the microcanonical
and canonical ensembles.  The equivalence and nonequivalence of the microcanonical
and Gaussian ensembles depends on the relationships between the sets $\eu$ and $\egbeta$
of corresponding equilibrium macrostates, which in turn are determined by support properties
of the generalized microcanonical entropy $s_\gamma(u) = s(u) - \gamma u^2$.  
As we just saw, for each $u \in N = (u_0,u_r)$, 
the microcanonical and canonical ensembles are nonequivalent.  
For $u \in N$ we would like to recover equivalence by 
replacing the canonical ensemble by an appropriate Gaussian ensemble.

Unfortunately, Theorem \ref{thm:applysigma1} is not applicable.
Although the first three of the hypotheses are valid, unfortunately
$s''$ is not bounded above on the interior of $\mboxdoms$.  Indeed, using the 
explicit formula for $s$ given in Theorem 4.3 in \cite{CosEllTou1}, one
verifies that $\lim_{u \goto (u_r)^-} s''(u) = \infty$.  
However, we can appeal to Theorem \ref{thm:localapply}, which is applicable
since $s$ is twice continuously differentiable on $N$.  We conclude that for each $u \in N$
and all sufficiently large $\gamma$ there exists a corresponding Gaussian ensemble
that is equivalent to the microcanonical ensemble for that $u$.   

By using other conjectured properties of the microcanonical entropy, we are able to deduce
the stronger result on the equivalence of the microcanonical and Gaussian ensembles
stated in Theorem \ref{thm:fgamma}.  As before, we denote
$\mboxdoms$ by $[u_\ell,u_r]$, where $u_\ell = -\frac{1}{2}$ and $u_r = -\frac{1}{2q}$, and write
\[
s'(u_\ell) = \lim_{u \goto (u_\ell)^+} s'(u) \ \mbox{ and } \ 
s'(u_r) = \lim_{u \goto (u_r)^-} s'(u)
\]
with a similar notation for $s''(u_\ell)$ and $s''(u_r)$. 
Using the explicit but complicated formula for $s$ given in Theorem 4.2 in 
\cite{CosEllTou1}, the following conjecture was verified numerically for
all $q \in \{4,5,\ldots,10^4\}$ and all $u \in (u_\ell,u_r)$ of the
form $u = u_\ell + 0.02 k$, where $k$ is a positive integer.

\begin{conj} 
\label{conj:sprimeprimeprime}
For all $q \geq 3$ the microcanonical entropy $s$ has the following two properties.

{\em (a)}  $s'''(u) > 0$ for all $u \in (u_\ell,u_r)$.

{\em (b)}  $s'(u_\ell) = \infty, \ 0 < s'(u_r) < \infty, \ s''(u_\ell) = - \infty, \
\mbox{ and } \ s''(u_r) = \infty$.  
\end{conj}

The conjecture implies that
$s''$ is an increasing bijection of $(u_\ell,u_r)$ onto $\R$.
Therefore, there exists a unique point $w_0 \in (u_\ell,u_r)$
such that $s''(u) < 0$ for all $u \in (u_\ell,w_0)$, $s''(w_0) = 0$,
and $s''(u) > 0$ for all $u \in (w_0,u_r)$.  It follows that the restriction
of $s$ to $[u_\ell,w_0]$ is strictly concave and the restriction 
of $s$ to $[w_0,u_r]$ is strictly convex.  These properties, which can 
be seen in Figure 1, are summarized
by saying that $s$ is a concave-convex function with break point $w_0$.

The interval $N = (u_0,u_r)$ exhibited in Figure 1
contains all energy values $u$ for which there exists no canonical 
ensemble that is equivalent with the microcanonical ensemble. 
Assuming the truth of Conjecture \ref{conj:sprimeprimeprime}, we now show that for each 
$u \in N$ there exists $\gamma \geq 0$ and an associated
 Gaussian ensemble that is equivalent with the microcanonical
ensemble for all $v \leq u$.  In order to do this, for $\gamma \geq 0$ 
we bring in the generalized microcanonical entropy
\[
s_\gamma(u) = s(u) - \gamma u^2
\]
and note that the properties of $s$ stated in Conjecture 
\ref{conj:sprimeprimeprime} are invariant
under the addition of the quadratic $-\gamma u^2$.  Hence,
if Conjecture \ref{conj:sprimeprimeprime} is valid, then $s_\gamma$ satisfies
the same properties as $s$.  In particular, $s_\gamma$ must be a concave-convex function
with some break point $w_\gamma$, which is the unique point in $(u_\ell,u_r)$
such that $s_\gamma''(u) < 0$ for all $u \in (u_\ell,w_\gamma)$, $s_\gamma''(w_\gamma) = 0$,
and $s_\gamma''(u) > 0$ for all $u \in (w_\gamma,u_r)$.  A straightforward argument, which 
we omit, and an appeal to Theorem \ref{thm:main}
show that there exists a unique point $u_\gamma \in (u_\ell,w_\gamma)$ having
the properties listed in the next three items.  These properties show
that $u_\gamma$ plays
the same role for ensemble equivalence involving the Gaussian ensemble that the point $u_0$ 
plays for ensemble equivalence involving the canonical ensemble.  

\begin{enumerate}
\item For $\gamma \geq 0$, $s_\gamma$ is strictly concave on the interval $(u_\ell, u_\gamma)$ and 
has a strictly supporting line at each $u \in (u_\ell, u_\gamma)$ and at $u_r$.  
Hence for $u \in F_\gamma = (u_\ell, u_\gamma) \cup \{u_r\}$ 
the ensembles are fully equivalent in the sense that there exists
$\beta$ such that $\eu = \egbeta$ [Thm.\ \ref{thm:main}(a)]. 
\item For $\gamma \geq 0$, $s_\gamma$ is concave but not strictly concave
at $u_\gamma$ and has a nonstrictly supporting line at $u_\gamma$ that also touches the graph
of $s$ over the right hand endpoint $u_r$.  Hence for $u \in  P_\gamma = \{u_\gamma\}$ the
ensembles are partially equivalent in the sense that there exists
$\beta$ such that $\eu \subset \egbeta$ but $\eu \not = \egbeta$
[Thm.\ \ref{thm:main}(b)].  
\item For $\gamma \geq 0$, 
$s_\gamma$ is not concave on the interval $N = (u_\gamma,u_r)$ and has no supporting line 
at any $u \in N$.  Hence for $u \in N_\gamma$ the ensembles are nonequivalent in the sense that 
for all $\beta$, 
$\eu \cap \egbeta = \emptyset$ [Thm.\ \ref{thm:main}(c)].
\end{enumerate}

We now state our main result.

\begin{thm}
\label{thm:fgamma}  We assume that Conjecture {\em \ref{conj:sprimeprimeprime}}
is valid.  Then as a function of $\gamma \geq 0$,
$F_\gamma = (u_\ell,u_\gamma) \cup \{u_r\}$ is strictly increasing, and as $\gamma \goto \infty$,
$F_\gamma \uparrow (u_\ell,u_r]$.  It follows that
for any $u \in N = (u_0,u_r)$, there exists $\gamma \geq 0$ such that the microcanonical ensemble
and the Gaussian ensemble defined in terms of this $\gamma$ 
are fully equivalent for all $v \in (u_\ell,u_r)$
satisfying $v \leq u$.  The value of $\beta$ defining the Gaussian ensemble
is unique and is given by $\beta = s'(v) - 2\gamma v$.
\end{thm}

The proof of the theorem relies on the next lemma, part (a) of which uses Proposition
\ref{prop:parabola}. When applied to $s_\gamma$, this proposition states that $s_\gamma$
has a strictly supporting line at a point if and only if $s$ has a strictly
supporting parabola at that point.  Proposition \ref{prop:parabola}
illustrates why one can achieve full 
equivalence with the Gaussian ensemble when full equivalence with the canonical ensemble fails. 
Namely, even when $s$ does not have a supporting line at a point, it might have a supporting
parabola at that point; in this case the supporting parabola can be made strictly supporting
by increasing $\gamma$.   The proofs of parts (b)--(d) of the next lemma rely on Theorem
\ref{thm:localapply} and on the properties
of the sets $F_\gamma$, $P_\gamma$, and $N_\gamma$ stated in the three 
items appearing just before the last theorem.

\begin{lemma}
\label{lem:moreproperties} 
We assume that Conjecture {\em \ref{conj:sprimeprimeprime}} is valid. Then
the following conclusions hold.

{\em (a)} If for some $\gamma \geq 0$, $s_\gamma$ has a supporting line at a point $u$, 
then for any $\tilde \gamma > \gamma$,
$s_{\tilde{\gamma}}$ has a strictly supporting line at $u$.

{\em (b)} For any $0 \leq \gamma < \tilde\gamma$, $F_\gamma \cup P_\gamma \subset F_{\tilde\gamma}$.

{\em (c)} $u_\gamma$ is a strictly increasing function of $\gamma \geq 0$ and $\lim_{\gamma \goto \infty}
u_\gamma = u_r$.  

{\em (d)} As a function of $\gamma \geq 0$, $F_\gamma$ is strictly increasing.
\end{lemma}

\noi
{\bf Proof.} 
(a)  Suppose that $s_\gamma$ has a supporting line at $u$ with slope $\bar\beta$. Then by Proposition
\ref{prop:parabola}
$s$ has a supporting parabola at $u$ with parameters $(\beta,\gamma)$, where $\beta = \bar\beta +
2 \gamma u$.  As the definition (\ref{eqn:parabola}) makes clear,
replacing $\gamma$ by any $\tilde\gamma > \gamma$ makes the supporting parabola
at $u$ strictly supporting.  Again by Proposition \ref{prop:parabola}
$s_{\tilde\gamma}$ has a strictly supporting line at $u$.  

(b) If $u \in F_\gamma \cup P_\gamma$, then $s_\gamma$ has a supporting line at $u$. 
Since $0 \leq \gamma < \tilde \gamma$, part (a) 
implies that $s_{\tilde \gamma}$ has a strictly supporting line at $u$.  Hence $u$
must be an element of $F_{\tilde \gamma}$.

(c) If $0 \leq \gamma < \tilde \gamma$, then by part (a) of the present lemma
$u_\gamma \in P_\gamma \subset F_{\tilde \gamma}$.  Since 
$F_{\tilde \gamma} = (u_\ell,u_{\tilde\gamma}) \cup \{u_r\}$ and 
since $u_\gamma < u_r$, it follows that $u_\gamma < u_{\tilde \gamma}$.   
Thus $u_\gamma$ is a strictly increasing function of $\gamma \geq 0$.
We now prove that $\lim_{\gamma \goto \infty} u_\gamma = u_r$.   For any $u \in (u_\ell,u_r)$,
part (b) of Theorem \ref{thm:localapply} states that
 there exists $\gamma_u > 0$ such that $s_{\gamma_u}(u)$ has a strictly supporting line at $u$.  
It follows that $u \in F_{\gamma_u} = (u_\ell,u_{\gamma_u}) \cup \{u_r\}$ and thus that
$u < u_{\gamma_u} < u_r$.  Since $u_\gamma$ is a strictly increasing function of $\gamma$, it follows that 
for all $\gamma > \gamma_u$, we have $u_\gamma > u_{\gamma_u} > u$.  We have shown that for any 
$u \in (u_\ell,u_r)$, there exists $\gamma_u > 0$ such that for all $\gamma >\gamma_u$, we have 
$u_\gamma > u$.  This completes the proof that $\lim_{\gamma \goto \infty} u_\gamma = u_r$.

(d)  Since $F_\gamma = (u_\ell,u_\gamma) \cup \{u_r\}$,
this follows immediately from the first property of $u_\gamma$ in part (c).  
The proof of the lemma is complete. \ \ink  

\skp

We are now ready to prove Theorem \ref{thm:fgamma}.  The properties
of $F_\gamma$ stated there follow immediately from Lemma \ref{lem:moreproperties}.
Indeed, since $u_\gamma$ is a strictly increasing function of $\gamma \geq 0$, 
$F_\gamma$ is also strictly increasing.  In addition, since $\lim_{\gamma \goto \infty} u_\gamma = u_r$
it follows that as $\gamma \goto \infty$, $F_\gamma \uparrow (u_\ell,u_r]$.
Since $F_\gamma$ is the set of full ensemble equivalence, we conclude that 
for any $u \in N = (u_0,u_r)$, there exists $\gamma > 0$ such that the microcanonical
ensemble and the Gaussian ensemble defined in terms of this $\gamma$ are fully equivalent for all $v \in (u_\ell,u_r)$
satisfying $v \leq u$.  The last statement concerning $\beta$ is 
a consequence of part (c) of Theorem \ref{thm:localapply}.  The proof of Theorem \ref{thm:fgamma}
is complete.

\begin{acknowledgments}
The research of Marius Costeniuc and Richard S.\ Ellis
was supported by a grant from the National Science Foundation (NSF-DMS-0202309), 
the research of Bruce Turkington was supported by a grant from the National Science
Foundation (NSF-DMS-0207064), and the research of Hugo Touchette was supported by
the Natural Sciences and Engineering Research Council of Canada and the Royal 
Society of London (Canada-UK Millennium Fellowship).
\end{acknowledgments}

\end{document}